%% file: EXO-12-037_temp.tex
\pdfoutput=1

\documentclass[11pt,twoside,a4paper,cmspaper,final,collab]{cms-tdr}
\def\svnVersion{282572}\def\cmsCernNo{2013-037}\def\cmsCernDate{\today}\def\cmsMessage{Published in Physical Review D as \href{http://dx.doi.org/10.1103/PhysRevD.91.052012}{\doi{10.1103/PhysRevD.91.052012.}}}

\begin{document}\cmsNoteHeader{EXO-12-037}

\hyphenation{had-ron-i-za-tion}
\hyphenation{cal-or-i-me-ter}
\hyphenation{de-vices}
\RCS$Revision: 281700 $
\RCS$HeadURL: svn+ssh://svn.cern.ch/reps/tdr2/papers/EXO-12-037/trunk/EXO-12-037.tex $
\RCS$Id: EXO-12-037.tex 281700 2015-03-21 22:43:05Z tomalini $
\ifthenelse{\boolean{cms@external}}{\providecommand{\CL}{C.L.\xspace}}{\providecommand{\CL}{CL\xspace}}
\newcommand{\BR}{\ensuremath{\mathcal{B}}\xspace}
\newcommand{\chidecay}{\ensuremath{\chiz\to\ell^+\ell^-\nu}\xspace}
\newcommand{\ctau}{\ensuremath{c\tau}\xspace}
\newcommand{\deltaPhi}{\ensuremath{\Delta\Phi}\xspace}
\newcommand{\dxysigma}{\ensuremath{\abs{d_0}/\sigma_d}\xspace}
\newcommand{\fb}{\unit{fb}}
\newcommand{\Higgs}{\ensuremath{\PH}\xspace}
\newcommand{\Higgsdecay}{\ensuremath{\PH\to\X\X}\xspace}
\newcommand{\piOverTwo}{\ensuremath{\pi/2}\xspace}
\newcommand{\pll}{\ensuremath{\overline{p}_{\ell\ell}}\xspace}
\newcommand{\sqasq}{\ensuremath{\PSQ\PSQ^*+\PSQ\PSQ}\xspace}
\newcommand{\sqdecay}{\ensuremath{\PSQ\to \cPq\chiz,~\chiz\to\ell^+\ell^-\nu}\xspace}
\newcommand{\squark}{\PSQ\xspace}
\newcommand{\vVec}{\ensuremath{\overline{v}_{\ell\ell}}\xspace}
\newcommand{\X}{\ensuremath{\cmsSymbolFace{X}}\xspace}
\newcommand{\Xdecay}{\ensuremath{\X\to\ell^+\ell^-}\xspace}
\newcommand{\Xdecayee}{\ensuremath{\X\to \EE}\xspace}
\newcommand{\Xdecaymumu}{\ensuremath{\X\to \MM}\xspace}

\cmsNoteHeader{EXO-12-037} 

\title{Search for long-lived particles that decay into final states containing two electrons or two muons
       in proton-proton collisions at \texorpdfstring{$\sqrt{s}=8$\TeV}{sqrt(s) = 8 TeV}}

\date{\today}

\abstract{
A search is performed for long-lived particles that decay into final states that include a pair of electrons
or a pair of muons. The experimental signature is a distinctive topology consisting of a pair of charged
leptons originating from a displaced secondary vertex. Events corresponding to an integrated luminosity of
19.6\,(20.5)\fbinv in the electron (muon) channel were collected with the CMS detector at the CERN LHC in
proton-proton collisions at $\sqrt{s} = 8$\TeV. No significant excess is observed above standard model
expectations. Upper limits on the product of the cross section and branching fraction of such a signal are
presented as a function of the long-lived particle's mean proper decay length. The limits are presented in an
approximately model-independent way, allowing them to be applied to a wide class of models yielding the above
topology. Over much of the investigated parameter space, the limits obtained are the most stringent to date.
In the specific case of a model in which a Higgs boson in the mass range 125--1000\GeVcc decays into a pair of
long-lived neutral bosons in the mass range 20--350\GeVcc, each of which can then decay to dileptons, the upper
limits obtained are typically in the range 0.2--10\fb for mean proper decay lengths of the long-lived
particles in the range 0.01--100\cm. In the case of the lowest Higgs mass considered (125\GeVcc), the limits
are in the range 2--50\fb. These limits are sensitive to Higgs boson branching fractions as low as $10^{-4}$.
}

\hypersetup{%
pdfauthor={CMS Collaboration},%
pdftitle={Search for long-lived particles that decay into final states containing two electrons or two muons in proton-proton collisions at sqrt(s) = 8 TeV},%
pdfsubject={CMS},%
pdfkeywords={CMS, physics}}

\maketitle

\section{Introduction}

Long-lived particles, which could manifest themselves through their delayed decays to leptons, are predicted
in many extensions of the standard model. For example, such particles could occur in:
supersymmetric (SUSY) scenarios such as ``split SUSY''~\cite{Hewett:2004nw} or SUSY with very weak
R-parity violation~\cite{Barbier:2004ez}, ``hidden valley'' models~\cite{Han:2007ae}, and the ``minimal $B-L$
extension of the standard model'' \cite{Basso:2008iv}.

In this paper we present an inclusive search for massive, long-lived exotic particles that decay to final
states that include a pair of charged leptons using proton-proton (\Pp\Pp) collision data collected at
$\sqrt{s}=8$\TeV during 2012 with the CMS detector at the CERN LHC.
Specifically, we search for events containing a pair of electrons or
muons (dileptons) originating from a common secondary vertex within the volume of the CMS tracker,
and with a significant transverse displacement from the event primary vertex. This topological signature
has the potential to provide clear evidence for physics beyond the standard model (SM). Furthermore, it
is almost free of background from SM processes.

The search results are formally obtained within the context of two specific models, however, they are presented in an
approximately model-independent way, allowing them to be applied to a wide range of models in which long-lived
particles decay to final states that include dileptons.
In the first model, the long-lived particle is a spinless boson \X, which has a
nonzero branching fraction to dileptons. The \X is pair-produced in the decay of a non-SM Higgs boson,
\Higgsdecay, \Xdecay~\cite{Strassler:2006ri}, where the Higgs boson is produced through gluon-gluon fusion and
$\ell$ represents either an electron or a muon. In the second model, the long-lived particle is a neutralino
\chiz which can decay via R-parity violating couplings into a neutrino and two charged
leptons~\cite{Barbier:2004ez,Allanach:2006st}. The neutralino is produced in events containing a pair of
squarks, where a squark can decay via the process \sqdecay. Both models predict up to two displaced dilepton
vertices per event in the CMS tracker volume, of which we only require one to be found. In this paper, we will
use {\it LL particle} to refer to any long-lived particle, such as the \X or \chiz particle considered in our
signal models.

The search presented here is an update of a previous CMS analysis that used a smaller data sample collected at
$\sqrt{s} =7$\TeV~\cite{DisplacedLeptons2011} in 2011. Improvements to the previous search include the higher
integrated luminosity collected in 2012, which increases the sensitivity of the search, and an improved
analysis strategy, which substantially broadens the range of signal models to which the analysis is
sensitive. The analysis complements two recent CMS publications: one searching for events that contain one
electron and one muon from LL particle decays~\cite{Khachatryan:2014mea}, and another that searches for LL
particles decaying to dijets~\cite{CMS-EXO-12-038}.

The \DZERO Collaboration has published the results of a search for leptons from non-prompt decays in its tracker
volume~\cite{Abazov:2006as,Abazov:2008zm}, performed at $\sqrt{s} =1.96$\TeV at the Fermilab Tevatron.
The ATLAS Collaboration has also performed related searches for
long-lived particles using different decay channels~\cite{ATLAS:2012av,Aad:2012zx},
or lower-mass LL particles \cite{Aad:2014yea}, compared to those considered in this paper.

\section{CMS detector}

The central feature of the CMS apparatus is a superconducting solenoid of 6\unit{m} internal diameter
providing an axial field of 3.8\unit{T}. Within the field volume are a silicon pixel and strip tracker, a
lead tungstate crystal electromagnetic calorimeter (ECAL), and a brass and scintillator hadron calorimeter
(HCAL). Muons are identified in gas-ionisation detectors embedded in the steel flux-return yoke of
the solenoid. A detailed description of the complete CMS detector,
together with a definition of the coordinate system used and the relevant kinematic variables,
can be found in Ref.~\cite{JINST}.

The silicon tracker is composed of pixel detectors (three barrel layers, and two forward disks at both ends of
the detector) surrounded by strip detectors (ten barrel layers, and three inner disks and nine forward disks
at both ends of the detector). The tracker covers the pseudorapidity range $\abs{\eta} < 2.5$. The pixel tracker
and a subset of the strip tracker layers provide three-dimensional measurements of hit positions. The other strip
tracker layers measure hit position only in ($r$, $\phi$) in the barrel, or ($z$, $\phi$) in the
endcap. Taking advantage of the strong magnetic field and the high granularity of the silicon tracker,
promptly produced charged particles with transverse momentum $\pt = 100$\GeVc are reconstructed with a
resolution of $\approx$1.5\% in \pt and of $\approx$15\mum in transverse impact parameter $d_0$.  The track
reconstruction algorithms~\cite{Chatrchyan:2014fea} are able to reconstruct displaced tracks with transverse
impact parameters up to $\approx$25\cm produced by particles decaying up to $\approx$50\cm from the beamline.
The performance of the track reconstruction algorithms has been studied with simulated events
~\cite{Chatrchyan:2014fea} and data~\cite{Khachatryan:2010pw}.  The silicon tracker is also used to reconstruct
the primary vertex position with a precision of 10--12\mum in each dimension.

The ECAL consists of nearly 76\,000 lead tungstate crystals, which provide coverage for $\abs{\eta} < 3$.
Its relative energy resolution improves with increasing energy. For energy deposits in the ECAL produced by
electrons or photons of $\ET\approx 60$\GeV, where $\ET=E\sin(\theta)$, the resolution varies between
1.1\% and 5\% depending on their pseudorapidity \cite{Chatrchyan:2013dga}.
Muons are measured in the range $\abs{\eta}< 2.4$ using detection planes
based on three technologies: drift tubes in the barrel region, cathode strip chambers in the endcaps,
and resistive-plate chambers in the barrel and endcaps.

The first level of the CMS trigger system, composed of custom hardware processors, selects events of interest
using information from the calorimeters and the muon detectors. A high-level trigger processor farm then
employs the full event information to further decrease the event rate.

\section{Data and simulated samples}
\label{sec:samples}

Data from pp collisions at $\sqrt{s}=8$\TeV, corresponding to an integrated luminosity of
$19.6 \pm 0.5$ ($20.5 \pm 0.5$)\fbinv, are used for the search in the electron (muon) channel. The lower
effective luminosity in the electron channel is due to different data quality requirements for the relevant
sub-detectors compared to those in the muon channel.

The electron channel data are collected with a high-level trigger~\cite{Khachatryan:2014ira}
that requires two clustered energy deposits in the ECAL. The leading (sub-leading)
energy deposit is required to have transverse energy $\ET > 36\,(22)$\GeV, and both clusters are required to
pass loose requirements on their compatibility with a photon/electron hypothesis. The muon channel
trigger requires two muons, each reconstructed in the muon detectors without using any primary vertex
constraint and having $\pt > 23$\GeVc. To suppress muons from cosmic rays, the
three-dimensional opening angle between the two muons must be less than 2.5\unit{radians}. Tracker information is
not used in either trigger, as the track reconstruction algorithm used in the high-level trigger (as opposed
to the standard offline track reconstruction) is not designed for finding displaced tracks.

For the \Higgsdecay model, simulated signal samples are generated using
{\PYTHIA~{v6.426}}~\cite{PYTHIA} to simulate \Higgs production through gluon-gluon fusion
($\cPg\cPg \to\Higgs$).  Subsequently, the \Higgs is forced to decay into $\X\X$, with the
\X bosons each decaying to dileptons (\Xdecay). Several samples are generated
with different combinations of the mass of the \Higgs ($m_{\Higgs}$ = 125, 200, 400, 1000\GeVcc) and the mass
of the \X boson ($m_{\X}$ = 20, 50, 150, 350\GeVcc). The Higgs boson resonance is assumed to be narrow
for the purposes of simulation, but the impact of this assumption on the analysis is negligible. Furthermore,
each sample is produced with three different \X boson lifetimes corresponding to mean
transverse decay lengths of approximately 2\cm, 20\cm, and 200\cm in the laboratory frame. For the \chidecay
model, \PYTHIA is used to simulate squark pair production and subsequent decay to \chiz, using four
combinations of squark and neutralino masses ($m_{\squark}, m_{\chiz}$) = (1500, 494), (1000, 148), (350, 148), and (120, 48)\GeVcc.  The R-parity violating couplings $\lambda_{122}$ and $\lambda_{121}$ are set to
nonzero values to enable the decay of the \chiz into two charged leptons and a neutrino. The values of
$\lambda_{122}$ and $\lambda_{121}$ are chosen to give a mean transverse decay length of approximately 20\cm. The
chosen masses explore the range to which CMS is currently sensitive.

Several simulated background samples are also generated with \PYTHIA.
The dominant background is Drell--Yan production of dileptons: prompt \EE or \MM pairs can be misidentified as
displaced from the primary vertex due to detector resolution effects, and the production and decay of \TT
pairs can produce genuinely displaced leptons, although the probability that both $\tau$-leptons decay
leptonically is small.
Other simulated backgrounds are from \ttbar, \PW/\Z boson pair
production (dibosons) with leptonic decays, and QCD multijet events. The last includes a potential
background source from semileptonic decays of \cPqb/\cPqc-flavour hadrons. In all samples, the response of
the detector is simulated using \GEANTfour~\cite{GEANT4}, and all the events are processed through the
trigger emulation and event reconstruction chains of the CMS experiment.

\section{Event reconstruction and selection}
\label{sec:RecoAndSelection}

To select pp collisions, events are required to contain a primary vertex with
at least four associated tracks and a position displaced from the nominal interaction point by no
more than 2\cm in the direction transverse to the beam, and no more than 24\cm in the direction
along the beam. Furthermore, to reject events produced by the interaction of beam-related protons
with the LHC collimators, for events with at least 10 tracks, the fraction of tracks classified as
``high purity'', as defined in Ref.~\cite{Chatrchyan:2014fea}, must exceed 25\%. When more
than one primary vertex is reconstructed in an event, we select the one with the largest sum of
the $\pt^{2}$ of the tracks associated to it.

In order to maximize the efficiency for reconstructing leptons from highly displaced vertices, we use lepton
identification algorithms that are less stringent than the standard CMS algorithms, which are not needed to
suppress the very low backgrounds in this analysis. Leptons are identified using tracks
reconstructed in the tracker that are classified as high purity, and have pseudorapidity $\abs{\eta}<2$.
The latter requirement is imposed because the efficiency for finding tracks from displaced secondary vertices decreases at large $\abs{\eta}$.

A track is identified as originating from an electron if its direction is consistent within a cone of size
$\Delta R = \sqrt{\smash[b]{(\Delta\eta)^2+(\Delta\phi)^2}} < 0.1$ with an energy deposit in the ECAL that is
reconstructed as a photon. Here, $\Delta\eta$ and $\Delta\phi$ are the differences between
the track and the energy deposit in the ECAL in $\eta$ and $\phi$, respectively. The energy of the electron is taken from the energy
deposit in the ECAL, since it is less affected by bremsstrahlung loss than is the measurement of the track \pt. Additional
quality requirements are placed on the ECAL energy deposit to reject background from hadronic sources.

A track is identified as originating from a muon if it matches a muon candidate found within
$\Delta R < 0.1$.  Here, $\Delta\eta$ and $\Delta\phi$ are the differences in direction between
the track and the muon found by the trigger in $\eta$ and $\phi$, respectively.

The LL particle candidates are formed from pairs of charged-lepton candidates. In the muon channel, the two
tracks must each have $\pt > 26$\GeVc and be oppositely charged. In the electron channel, the higher\,(lower)
\ET electron must satisfy $\ET > 40$\GeV\,(25\GeV). These thresholds are set slightly higher than the
corresponding trigger requirements to ensure that the selected events have high trigger efficiency. In the dielectron channel,
the two tracks must also satisfy $\pt > 36$\GeVc\,(21\GeVc) if associated to the higher\,(lower) \ET
electron. This \pt requirement, which is slightly lower than the corresponding \ET requirement placed on the
ECAL energy deposit, suppresses electrons that emit large amounts of bremsstrahlung, and which thus tend to have poor
impact parameter resolution. No charge requirement is applied to electrons, as the probability
of mismeasuring the charge is nonnegligible for high-\pt electrons.

To reject promptly produced particles, the tracks are required to have a transverse impact parameter significance with
respect to the primary vertex of $\dxysigma > 12$, where $\sigma_d$ is the uncertainty on $\abs{d_0}$. This value
is chosen to give an expected background significantly below one event, which gives the best signal sensitivity
for the vast majority of the LL particle lifetimes considered in this paper. Both lepton candidates are
required to be isolated, to reject background from jets. Specifically, a hollow isolation cone is constructed
around each candidate, with a radius $0.04 < \Delta R < 0.3$ for electrons and $0.03 < \Delta R < 0.3$ for muons.
Within this isolation cone, the ratio of the scalar $\sum\pt$ of all tracks with $\pt > 1$\GeVc, excluding the other
lepton candidate, to the \pt of the lepton, must be less than 0.1.

The two tracks are fitted to a common vertex, which is required to have $\chi^2/\mathrm{dof} < 10$\,(5) in the
electron\,(muon) channel.
To ensure that the candidate tracks were produced at this vertex, we require that the 
number of hits, between the centre of CMS and the vertex position, that are assigned 
to the tracks is no more than 1, and that the number of
missing hits on the tracks between the vertex position and the outer envelope of the tracker is no more than 3\,(4) in the
electron\,(muon) channel, where in both cases the numbers are summed over both tracks.
A missing hit is defined as occurring when a track passes through an active sensor without being assigned a
reconstructed hit. To eliminate background from \JPsi and $\Upsilon$ decays, and from $\gamma$ conversions, LL particle
candidates are required to have a dilepton invariant mass larger than 15\GeVcc.

Cosmic ray muons may be reconstructed as back-to-back tracks. To reject them, the three-dimensional opening
angle between the two muons must be less than 2.48~radians. This requirement is slightly tighter than the
requirement in the trigger. Background from misidentified leptons is reduced by requiring that the two lepton
candidates are not both matched to the same trigger object or offline photon. Owing to the difficulty of
modelling the low trigger efficiency for closely spaced muon pairs, the two muons are required to be separated
by $\Delta R > 0.2$.

Finally, the signed difference in azimuthal angles, $\deltaPhi$, between the dilepton momentum vector,
\pll, and the vector from the primary vertex to the dilepton vertex, \vVec, is required to satisfy
$\abs{\deltaPhi} <\piOverTwo$, where $\deltaPhi$ is measured in the range $0 < \deltaPhi < \pi$.
Dilepton candidates satisfying all other selection requirements, but with $\abs{\deltaPhi} > \piOverTwo$,
are used to define a control region, as detailed in Section~\ref{sec:bkgndest}.

Events containing at least one LL particle candidate that passes all selection requirements are accepted.
Where more than one candidate is found in an event, the one with largest \dxysigma is chosen.  The \dxysigma of a candidate is defined as the minimum of the two \dxysigma values of the leptons that comprises it.

The overall signal efficiency is defined as the fraction of events in which at least one dilepton candidate
passes all selection criteria. It is determined from the simulated signal samples, separately for the electron and muon
channels, and independently for two different classes of events: first for events in which only one LL
particle (\X or \chiz) decays to the chosen lepton species, defining efficiency $\epsilon_{1}$, and second for
events in which both LL particles decay to the chosen lepton species, defining efficiency
$\epsilon_{2}$. The efficiencies are estimated for LL particle lifetimes corresponding to mean
transverse decay lengths in the range of 200\mum--200\unit{m}, by reweighting the simulated signal events. The maximum
value of $\epsilon_{1}$, which is attained for \Higgsdecay with $m_{\Higgs} = 1000$\GeVcc,
$m_X = 150$\GeVcc, and $c\tau = 1$\cm, is approximately 36\%\,(46\%) in the electron\,(muon) channel, but it
becomes significantly smaller at lower \Higgs masses or at longer and shorter lifetimes. For example, if $c\tau$ is
increased to 20\cm for this set of masses, then $\epsilon_{1}$ drops to 14\%\,(20\%) in the electron\,(muon)
channel. The efficiencies in the muon channel are generally higher because of the lower \pt thresholds compared to
the corresponding thresholds in the electron channel.

In order to reduce the model dependence of our results, it is useful to define a set of acceptance criteria
that specify the LL particles decaying to dilepton final states that can be reconstructed in the CMS detector.
Specifically, the generated transverse decay length of the LL particle should be no more than 50\cm, and the
generated electrons\,(muons) should satisfy the same \ET\,(\pt) and $\eta$ requirements that are applied to the
reconstructed electrons\,(muons), which are listed earlier in this section.  The acceptance $A$ is defined as
the fraction of LL particle decays that pass the acceptance criteria.  Re-evaluating the signal efficiency
$\epsilon_{1}$, using only LL particle decays within the acceptance, yields $\epsilon_{1}/A$, which is larger
than $\epsilon_{1}$.  For example, for $m_{\Higgs} = 1000$\GeVcc, $m_X = 150$\GeVcc, and $c\tau =1$\cm, the
value of $\epsilon_{1}/A$ is approximately 44\%\,(58\%) in the electron\,(muon) channel. More importantly, the
efficiency defined in this way shows much less dependence on the choice of signal model; \eg for this same
choice of masses, but with $c\tau = 20$\cm, it falls only to 28\%\,(40\%) in the electron\,(muon) channel.

\section{Background estimation and associated systematic uncertainties}
\label{sec:bkgndest}

To estimate the background, we consider the quantities \vVec, \pll and \deltaPhi defined in Section~\ref{sec:RecoAndSelection}.
For signal events, \vVec corresponds to the flight direction of the LL particle, and assuming that the dilepton system produced when
the LL particle decays is usually boosted with respect to its flight direction, 
the direction of \pll is correlated with that of \vVec.
In contrast, for background events, \vVec does not correspond to the flight direction of any long-lived particle, so its angular
distribution with respect to \pll should not show any forward-backward asymmetry.
For example, in the case of Drell--Yan production of $\ell^+\ell^-$, \vVec is determined only by effects such as detector resolution
or primary vertex misassignment. Although in the case of Drell--Yan production of \TT, leptonic products of the $\tau$-lepton
decays may have significant values of $\dxysigma$ because of the nonzero lifetime of the $\tau$-lepton, a vertex reconstructed
from two such leptons would not correspond to a genuine particle decay vertex.
Processes such as non-prompt \JPsi decay or $\gamma$-conversions, which can give rise to genuine displaced dilepton vertices,
are eliminated by the requirement on the minimum dilepton mass. Cosmic ray background is reduced to negligible levels
via the dimuon opening angle requirement that rejects back-to-back muons.

Therefore if we define a signal region with $\abs{\deltaPhi} < \piOverTwo$ and a control region with
$\abs{\deltaPhi} > \piOverTwo$, we expect that signal events will populate the former region, while
background events will be equally distributed between the two. Consequently, we can use the distribution of
events in the control region to derive a data-driven estimate of the background expected in the signal
region.

\begin{figure*}[hbtp]
  \begin{center}
    \includegraphics[width=0.49\textwidth]{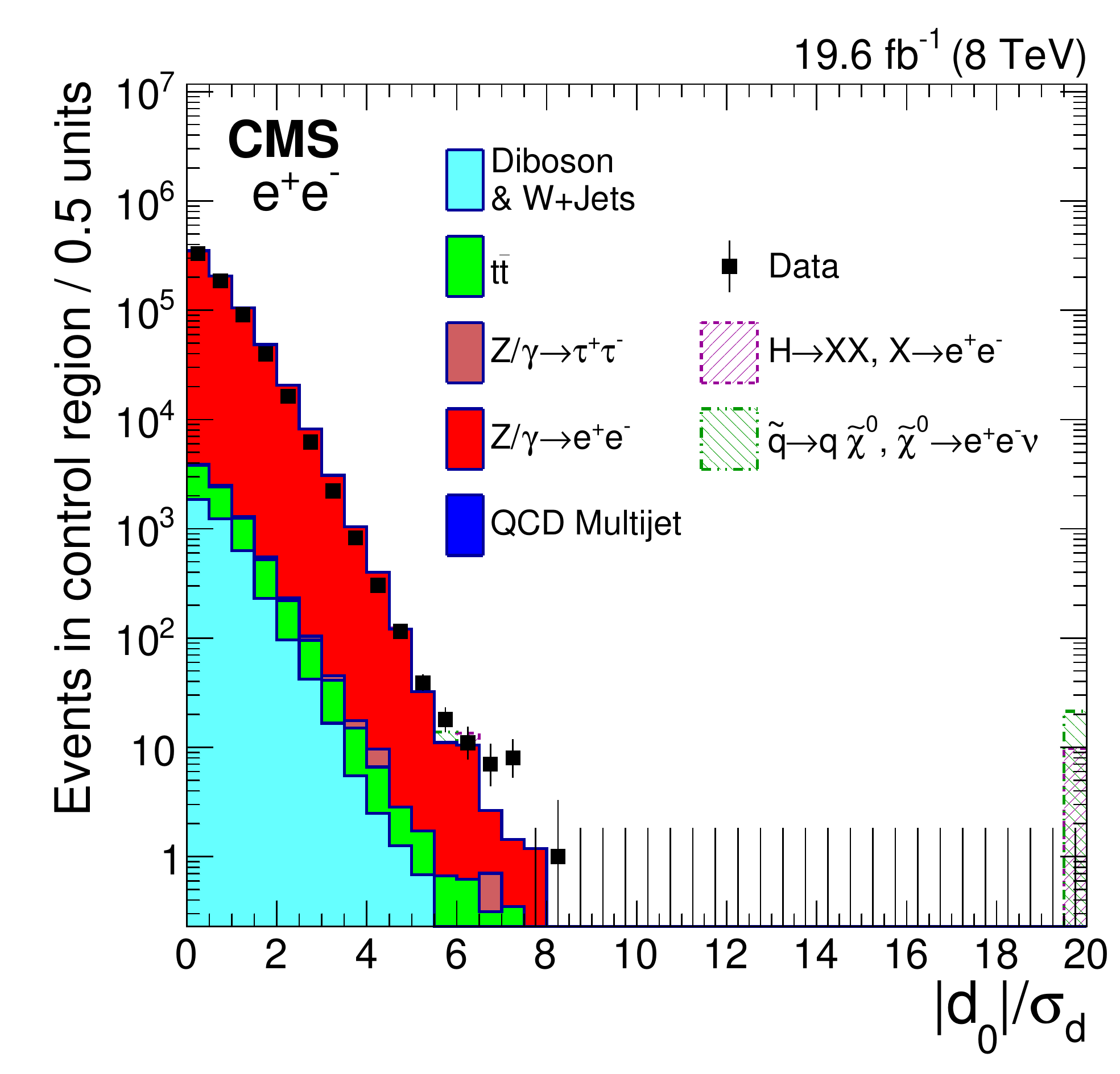}
    \includegraphics[width=0.49\textwidth]{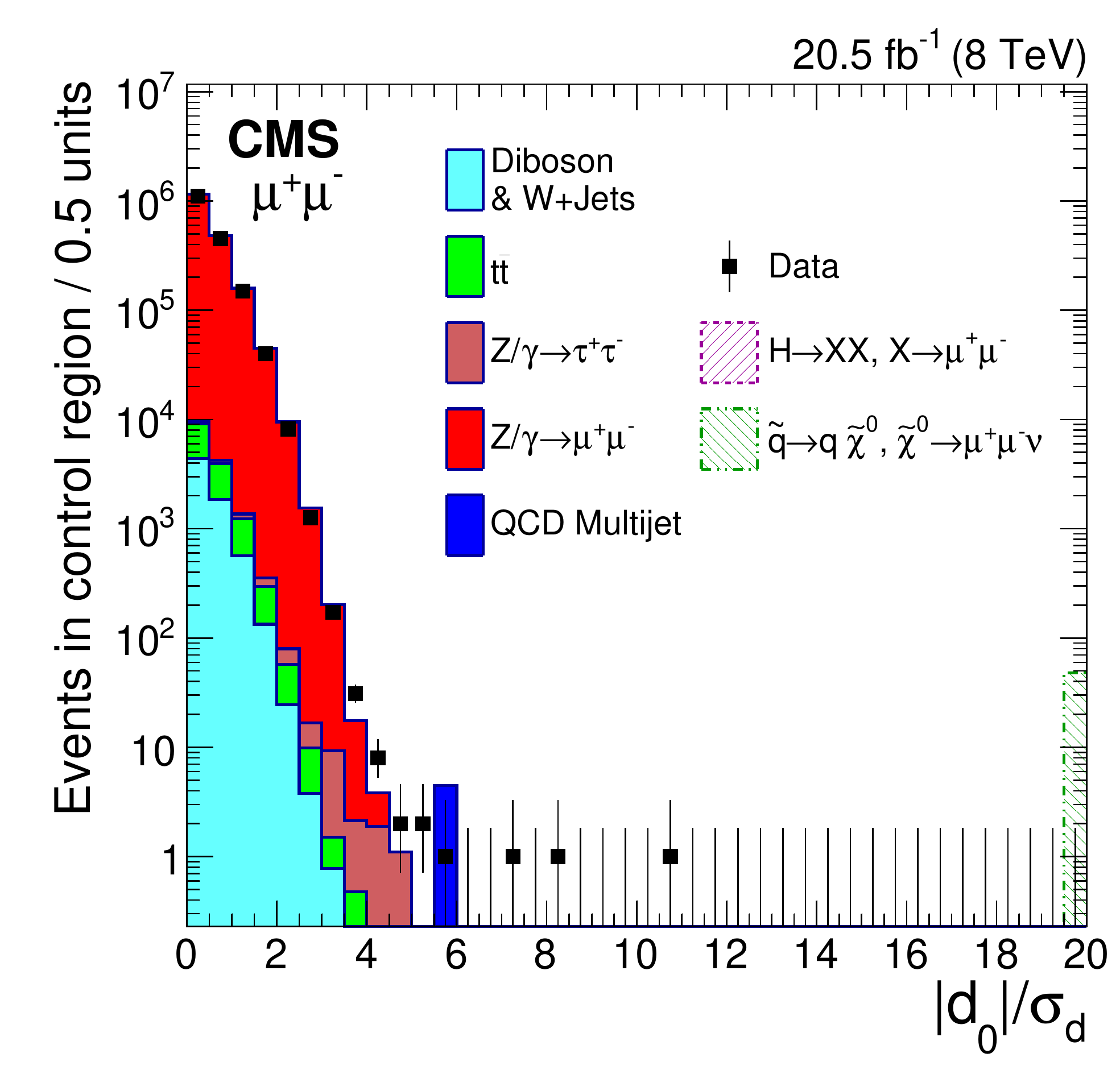}
    \includegraphics[width=0.49\textwidth]{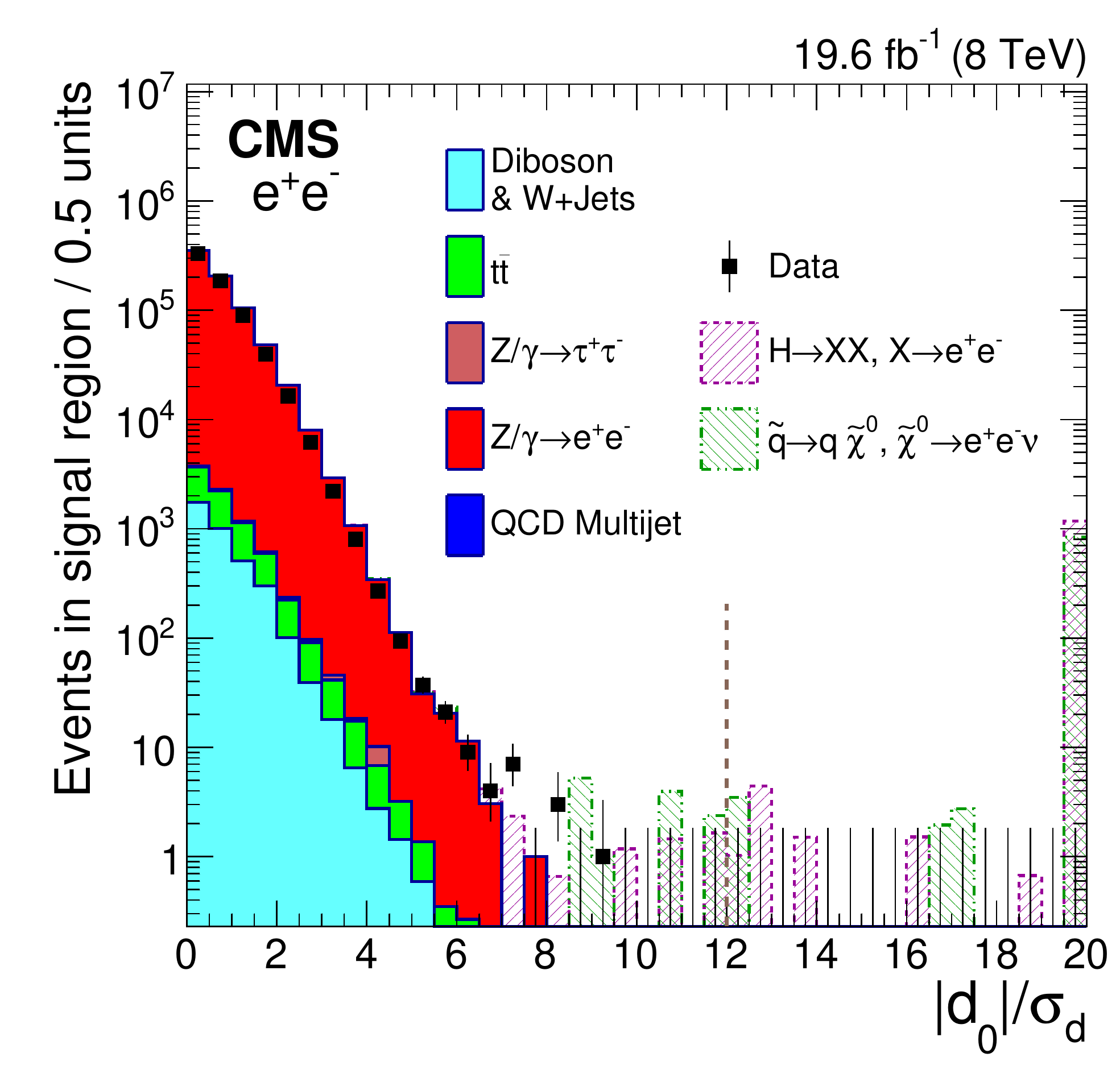}
    \includegraphics[width=0.49\textwidth]{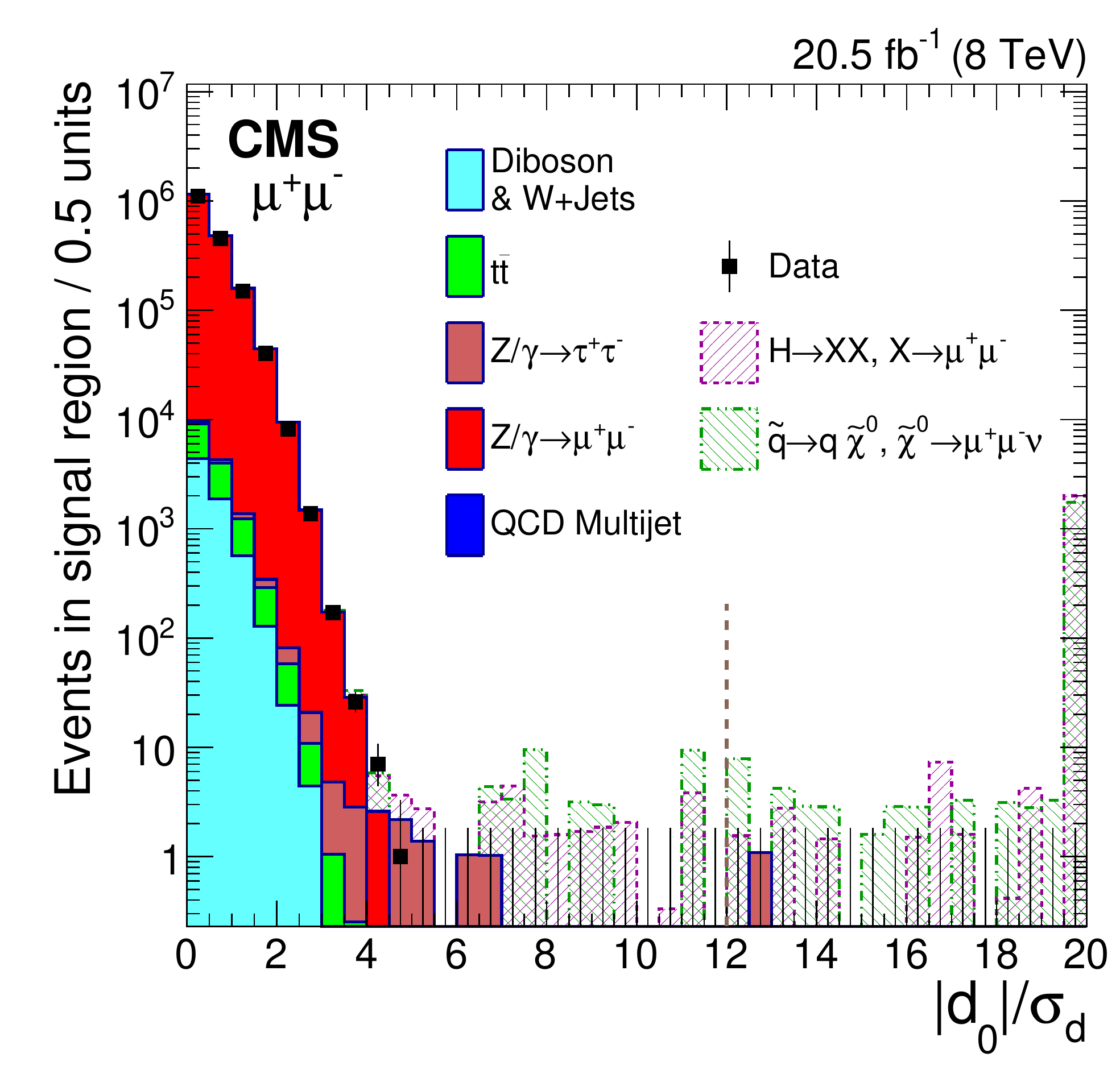}
    \caption{The \dxysigma distribution for the electron\,(left) and muon\,(right) channels, shown
    in the top row for events in the control region ($\abs{\deltaPhi} > \piOverTwo$) and in the bottom row for
    events in the signal region ($\abs{\deltaPhi} < \piOverTwo$).
    Of the two leptons forming a candidate, the distribution of the one with the smallest \dxysigma is plotted.
    The solid points indicate the data, the shaded histograms are the simulated
    background, and the hashed histograms show the simulated signal.
    The histogram corresponding to the \Higgsdecay model is shown for $m_{\Higgs} = 1000$\GeVcc and
    $m_X = 350$\GeVcc. The histogram corresponding to the \chidecay model is shown for $m_{\squark} = 350$
   \GeVcc and $m_{\chiz} = 140$\GeVcc.
    The background histograms are stacked, and each simulated signal sample is independently stacked on top of the total simulated background.
    The $d_0$ corrections for residual tracker misalignment, discussed in the text, have been applied.
    The vertical dashed line shows the selection requirement $\dxysigma > 12$.
    Any entries beyond the right-hand side of a histogram are shown in the last visible bin of the histogram.}
    \label{fig:D0SigmaDataMC}
  \end{center}
\end{figure*}

Figure~\ref{fig:D0SigmaDataMC} shows the \dxysigma distribution of the simulated events in
the signal and control regions. Each of the simulated backgrounds is statistically consistent with being symmetrically divided
between the two regions. The expected background is predominantly Drell--Yan dilepton production, with some
contribution from QCD multijets. Any discrepancies between data and simulation are unimportant since the analysis
uses a data-driven background estimate. They may arise because of imperfect modelling in the simulation or because of the
large statistical uncertainty in the simulated QCD multijet background. The multijet background near
$\dxysigma=6$ in the top, right-hand plot corresponds to a single simulated event.  We observe that more than 97\%\,(95\%)
of simulated signal events fall into the signal region for the $\Xdecay$\,($\chidecay$) model for all the samples considered.

\begin{figure*}[tpb]
  \begin{center}
    \includegraphics[width=0.49\textwidth]{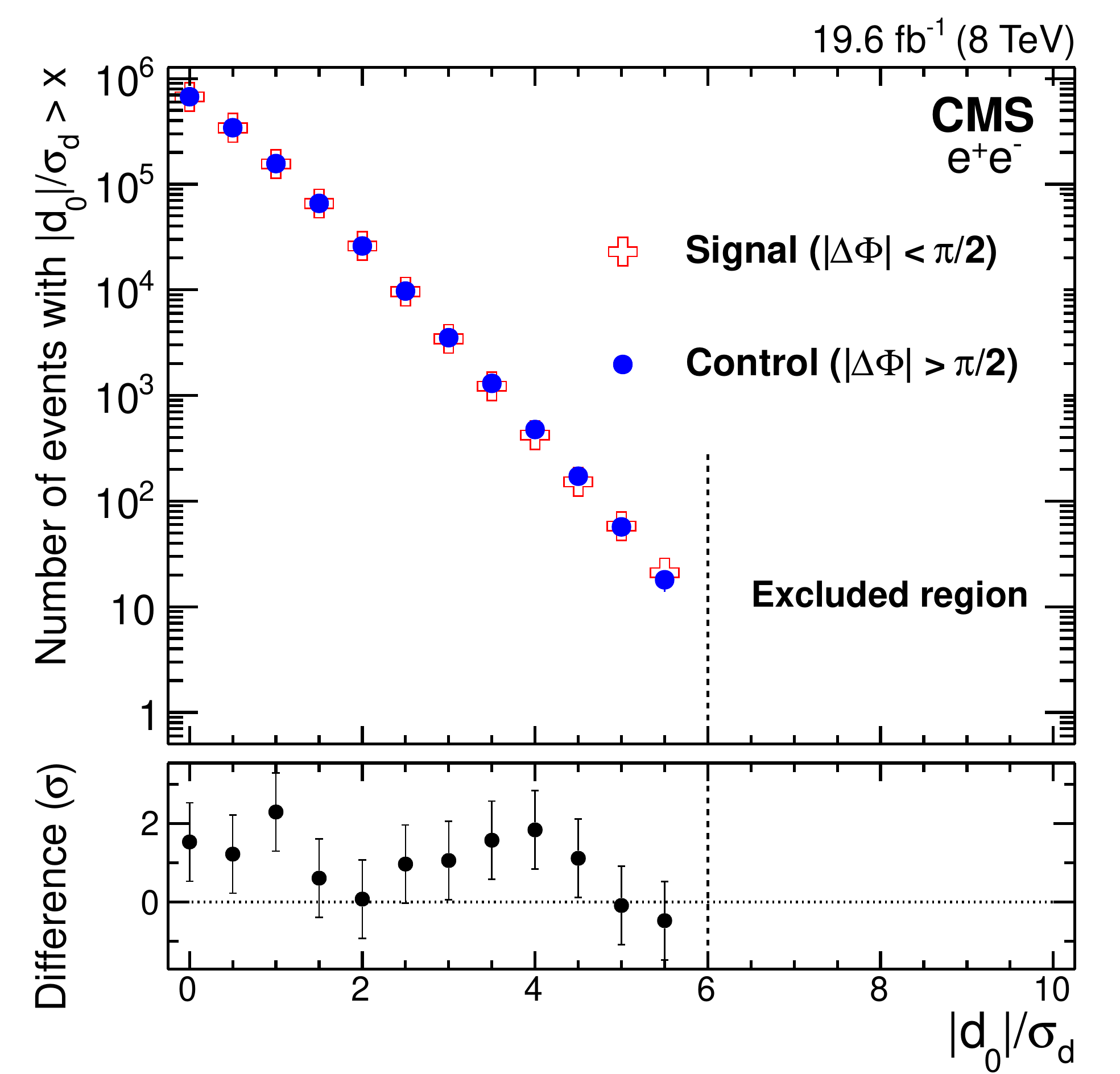}
    \includegraphics[width=0.49\textwidth]{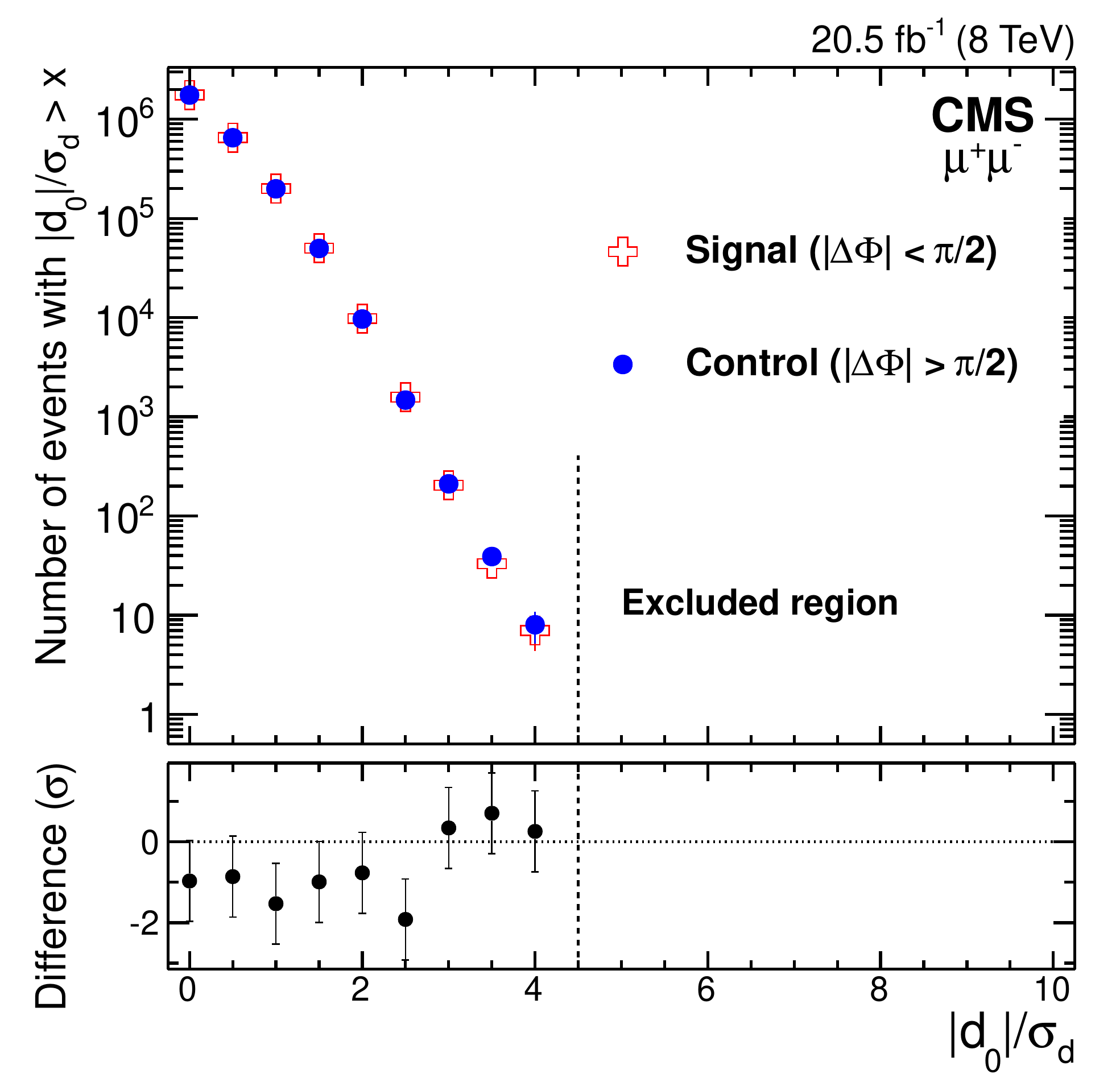}
    \caption{Comparison of the tail-cumulative distributions of \dxysigma for data in the signal region
      ($\abs{\deltaPhi} < \piOverTwo$) and the control region ($\abs{\deltaPhi} > \piOverTwo$) for the electron
      channel (left) and the muon channel (right). The $d_0$ corrections for residual tracker misalignment,
      discussed in the text, have been applied. Of the two leptons forming a candidate, the distribution of the one with the
      smallest \dxysigma is plotted. The bottom panels show the statistical significance
      of the difference between the distributions in the signal and control regions.}
    \label{fig:FinalCorrections}
  \end{center}
\end{figure*}

Besides using simulated events, we validate this method by comparing the $\dxysigma$ distribution in the
signal region with the one in the control region using data at $\dxysigma$ values for which the
sample is background-dominated. Figure~\ref{fig:FinalCorrections} shows the tail-cumulative distributions,
which are defined as integrals from the plotted value to infinity, of \dxysigma in the signal and control regions.
However, the region with $\dxysigma > 6$\,(4.5) in the electron\,(muon) channel is excluded from the integral,
to ensure that the signal region is background-dominated.  No statistically significant difference between the
two regions is seen.

We observe zero events in data with $\dxysigma > 12$ in the control region, and this determines the
probability distribution of the expected background level, as discussed in Section~\ref{sec:results}. The
systematic uncertainty in this estimate is defined below.

Residual misalignment of the tracker is the only effect that can cause the expected background to differ
significantly in the signal and control regions. This effect is largely removed by applying corrections, described below, to the
conventionally signed \cite{Chatrchyan:2014fea} transverse and longitudinal ($z_0$) impact parameters of all tracks. The
mean offset from zero of the signed $d_0$ and $z_0$ of prompt muon tracks (\ie $\abs{d_0}$ and $\abs{z_0}$ below 500\mum)
is measured as a function of the track $\eta$ and $\phi$, and also as a function of run period. This
bias, which arises from residual misalignment and is always less than 5\mum, is then subtracted from the
measured impact parameters of individual tracks. To verify that this method is reliable, we first apply it to a data
sample reconstructed with a preliminary alignment calibration, much inferior to the final alignment
calibration used for the latest CMS datasets. In this sample, we observe a significant asymmetry between the
control and signal regions, most of which disappears when the impact parameter corrections are
applied.

Two approaches, described below, are used to assess the effect of any remaining systematic uncertainty in the background estimate due to misalignment. The first makes a direct measurement of the background asymmetry in the \dxysigma distribution.  The second checks how much, if at all, the LL particle search results change if the impact parameter corrections are removed.

The first approach measures the systematic uncertainty remaining after the impact parameter corrections have
been applied, by comparing the \dxysigma distributions in the two regions with $\deltaPhi < 0$ and $\deltaPhi >
0$.  Both signal and background are expected to be equally divided between these two regions, so any
significant asymmetry between them can only arise through systematic effects. We measure the size of this
asymmetry by comparing the ratio of the number of events in the tail-cumulative distribution of \dxysigma in
the region $\deltaPhi < 0$ with that in the region $\deltaPhi > 0$. Points at \dxysigma values with very few
events, such that the relative statistical uncertainty in this ratio is greater than 30\%, are excluded since they
would not provide a precise estimate of the systematic uncertainty. The maximum difference of the
ratio from unity for all remaining points is then taken to be the systematic uncertainty. Using this procedure, we
obtain a systematic uncertainty of 11\% and 21\% in the electron and muon channels, respectively, in the
estimated amount of background.

The second approach addresses a potential issue with the first method, namely that it measures the
systematic uncertainty in the background normalization at lower values of \dxysigma than are used in our
standard selection.  In the data, the bias on the track $d_0$ due to misalignment is less than
5\mum, whereas our $\dxysigma > 12$ requirement typically corresponds to a selection on $\abs{d_0}$ of approximately
180\mum. This suggests that misalignment should not be a significant effect at large \dxysigma. Nonetheless,
to allow for the possibility that it might be, we employ the second approach; namely, when computing our final
limits, we do so twice, once with the impact parameter corrections applied, and once without them, and then
take the worse limits as our final result. This should be conservative, given that as stated above, the
impact parameter corrections remove the majority of any asymmetry caused by misalignment.
In practice, the misalignment is so small that these two sets of limits are identical.

\section{Systematic uncertainties affecting the signal}
\label{sec:Systematics}

The systematic effects influencing the signal
efficiency arise from: uncertainties in the efficiency of reconstructing tracks from displaced vertices, the
trigger efficiency, the modelling of pileup (\ie additional pp collisions in the same bunch crossing),
the parton distribution function (PDF) sets, the renormalization and factorization scales used in
generating simulated events, and the effect of higher-order QCD corrections.

Table~\ref{tab:syst} summarizes the nonnegligible sources of systematic uncertainty affecting the signal
efficiency.  These are discussed in more detail below. The most important sources are those related to the
track reconstruction efficiency. The relative uncertainty in the measurement of the integrated luminosity is
2.6\%~\cite{cms:lumi13}.

\begin{table*}[hbtp]
\centering
\topcaption{Systematic uncertainties affecting the signal efficiency over the two signal models and all mass
  values considered. In all cases, the uncertainty specified is a relative uncertainty. The NLO uncertainty is
  significant only for the $\Higgsdecay$ model with $m_{\Higgs} = 125$\GeVcc. The relative uncertainty in the
  integrated luminosity is 2.6\%.}
\label{tab:syst}
\begin{scotch}{ll}
Source & Uncertainty \\
\hline
Pileup modelling & 2\% \\
Parton distribution functions & ${<}1\%$ \\
Renormalization and factorization scales & ${<}0.5\%$ \\
Track reconstruction efficiency from cosmic ray muons & 6.1\% \\
Track reconstruction efficiency in high hit occupancy environment & 3.5\% \\
Track reconstruction efficiency loss due to bremsstrahlung (\Pe\ only)& 5.8\% \\
Trigger efficiency & 1.7\% (\Pe), 6.2\% ($\mu$) \\
NLO effects (only for the $m_{\Higgs} = 125$\GeVcc case) & 5--7\% \\
\end{scotch}
\end{table*}

Varying the modelling of the pileup within its estimated uncertainties yields a relative change in the signal
selection efficiency of less than 2\%, irrespective of the mass point chosen. The relative uncertainty due to
the choice of PDF set is studied using the PDF4LHC prescription~\cite{Bourilkov:2006cj} and is less
than 1\% for all mass points. The dependence of the acceptance on the choice of the renormalization and
factorization scales, which are chosen to be equal, is found to be well below 0.5\% when they are
varied by a factor of 0.5 or 2.  These uncertainties are applied in the cross section limit calculation.

\subsection{Track finding efficiency}

\begin{figure*}[hbtp]
  \centering
  \includegraphics[width =0.49\textwidth]{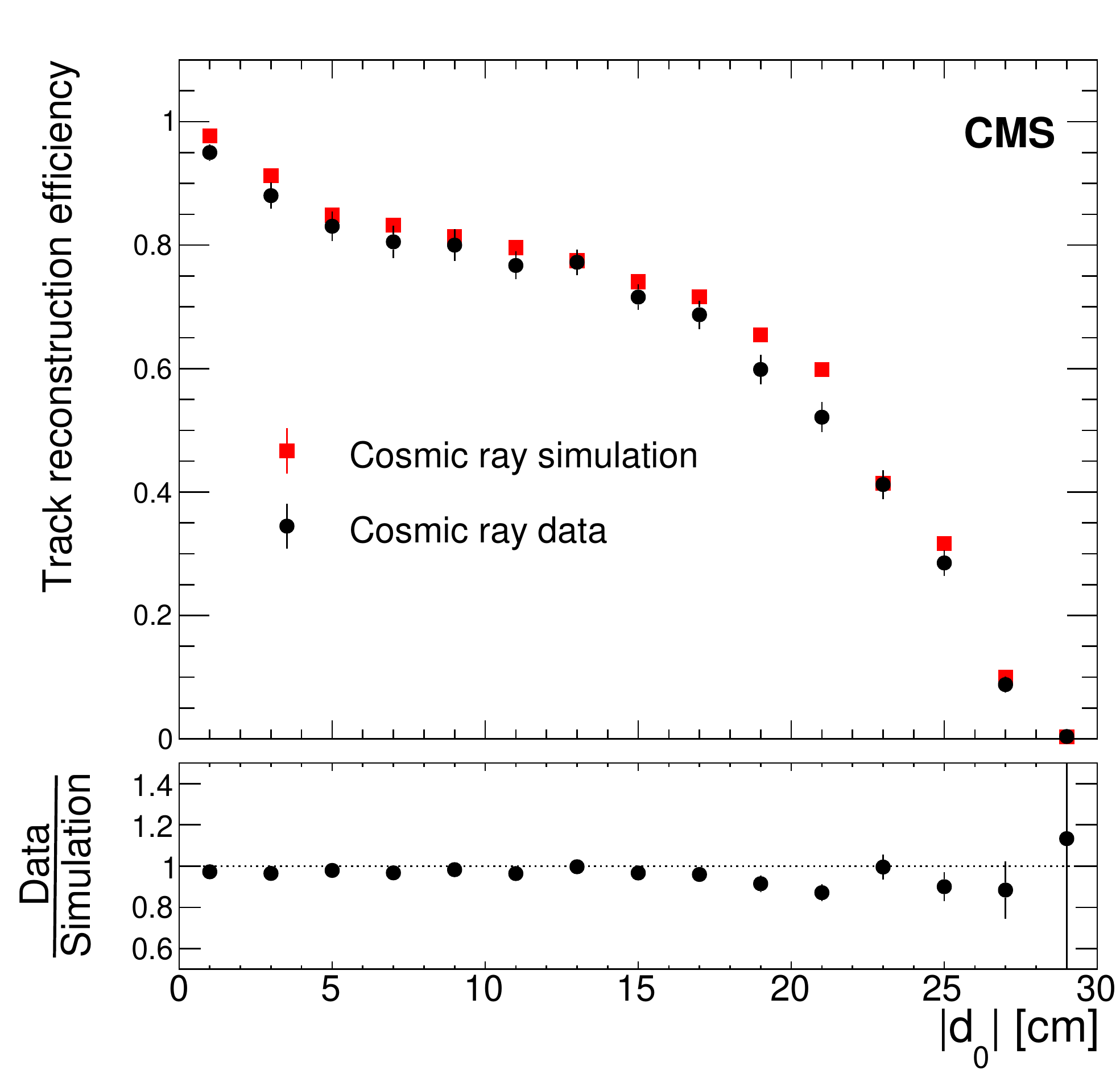}
  \includegraphics[width =0.49\textwidth]{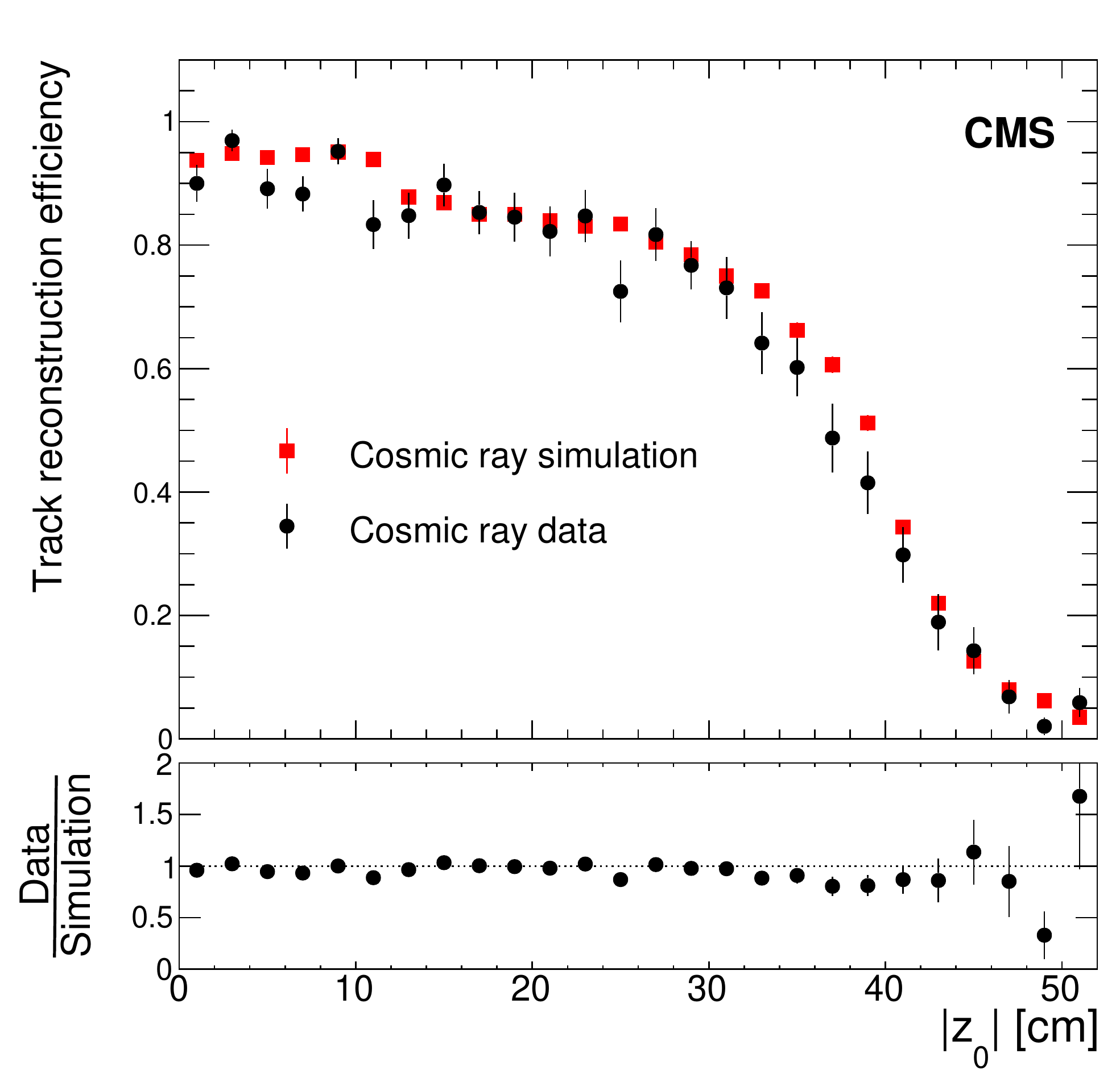}
  \caption{Efficiency to find a track in the tracker, measured using cosmic ray muons reconstructed in the muon
    detectors, as a function of the transverse (left) and longitudinal (right) impact parameters (relative
    to the nominal interaction point of CMS). The efficiency is plotted in bins of 2\cm width.  For the left
    plot, the longitudinal impact parameter $\abs{z_0}$ is required to be less than 10\cm, and for the right
    plot, the transverse impact parameter $\abs{d_0}$ must be less than 4\cm. The bottom panels show the ratio of
    the efficiency in data to that in simulation. The uncertainties in the simulation are smaller than the size
    of the markers and are not visible.}
  \label{fig:EffVsImpact}
\end{figure*}

Three methods are used to assess if the efficiency to reconstruct displaced tracks is correctly modelled by
the simulation. The first method consists of a direct measurement of the efficiency to reconstruct isolated,
displaced tracks, using cosmic ray muons. Events are selected from dedicated running periods with no beam present, and
the cosmic ray muons are reconstructed by combining the hits in the muon detectors from opposite halves of the
CMS detector. The efficiency to reconstruct, in the tracker, a track associated with a cosmic ray muon, as a function of the
transverse and longitudinal impact parameters, is shown in Fig.~\ref{fig:EffVsImpact}. The systematic uncertainty
on the dilepton efficiency is estimated as follows. We use the measured track reconstruction efficiency to estimate
the efficiency to reconstruct a pair of leptons of given impact parameters. We then weight this efficiency according
to the impact parameter distributions of the dileptons in the simulated signal Monte Carlo samples.
The ratio of the estimated efficiency per dilepton candidate in data to simulation differs from unity by no more
than 6.1\% for any of the samples considered, so this value is taken as the systematic uncertainty.

A second method is used to study how the presence of a high density of tracker hits around displaced
leptons degrades the track reconstruction performance. This method takes cosmic ray muon data, where each
muon is reconstructed in the muon detectors and is successfully associated to a track reconstructed
in the tracker. It embeds each of these tracks and its associated hits into a high-occupancy pp
collision data event, and measures the fraction of these embedded tracks that can still be
successfully reconstructed in this environment as a function of their impact parameters.
The results are compared with those obtained by embedding tracks from simulated cosmic events in
simulated pp collisions. The same procedure described at the end of the preceding paragraph is applied,
and leads us to conclude that the efficiency per candidate has an additional systematic
uncertainty, related to the track reconstruction efficiency in a high hit density environment, of 3.5\%.

A third method~\cite{CMS-EXO-12-038} uses charged pions from $\PKzS$ decay to establish that the track reconstruction
efficiency is simulated with a relative systematic uncertainty of 5\%. Since this method is mainly sensitive
to the track reconstruction efficiency of low-\pt hadrons in jets, it is used only to provide additional reassurance that
the displaced track reconstruction efficiency is well modelled.

These methods do not explicitly measure the track reconstruction efficiency for electrons, where
an additional systematic uncertainty must be considered. For the leptons from LL particle decay in the
simulated signal samples, the track reconstruction efficiency for the electrons is about 78\% that of the
muons, where the difference arises from the emission of bremsstrahlung. This difference does not show
significant variation with respect to the transverse decay length of the LL particle. The material budget
of the tracker is modelled in simulation to an accuracy of $<$10\%~\cite{pas-trk-10-003}. Since the amount
of bremsstrahlung should be proportional to the amount of material in
the tracker, this implies a corresponding relative uncertainty in the difference between the track
reconstruction efficiencies for electrons and muons. This leads to a bremsstrahlung-related
relative uncertainty in the tracking efficiency for electrons of $0.22\times10\%/(1-0.22) = 2.9$\%,
where the denominator arises because this uncertainty is measured relative to the tracking efficiency for electrons, not muons.
The corresponding systematic uncertainty for the dielectron candidates, which have two tracks, is twice as large, namely 5.8\%.

\subsection{Trigger efficiency}
\label{sec:trigEffi}

The trigger efficiency is measured using the ``tag-and-probe'' method~\cite{Khachatryan:2010xn}. In
the muon channel, \Z~boson decays to dimuons are reconstructed in data collected with
single-muon triggers. They are then used to measure the efficiency for a muon to pass the
selection criteria of one leg of the dimuon trigger used in this analysis. The dimuon trigger
efficiency is then obtained as the square of this single-muon efficiency, which assumes that there
is no correlation in efficiency between the two leptons. This is generally a good assumption except
for dimuons separated by $\Delta R < 0.2$, which are excluded because the trigger is inefficient for
closely spaced dimuons. In the electron channel, the method is similar, but since the two legs
of the trigger for this channel have different \ET thresholds, the efficiency of each leg is measured separately.
In data, the trigger efficiency is essentially 100\% for electrons satisfying the analysis selection.
Under the same conditions, the efficiency for muons with a \pt of about 26\GeVc is above 70\% and it reaches
a plateau of approximately 85\% for $\pt > 40$\GeVc.

The systematic uncertainty associated with the trigger efficiency is evaluated by taking the
difference between the efficiency estimates from data and simulation, which yields a total
relative uncertainty of 1.7\% for the electron channel and 6.2\% for the muon channel. To ensure
that the trigger efficiencies obtained from the sample of \Z~bosons, in which the leptons are
prompt, are also valid for leptons from LL particle decay, we examine the trigger efficiency in
simulated signal events as a function of the lifetime of the LL particles. For LL particles
passing the acceptance criteria defined in Section~\ref{sec:RecoAndSelection}, no statistically
significant dependence of the trigger efficiency on their lifetime is seen. Therefore,
systematic uncertainties related to this source may be neglected in
comparison to the systematic uncertainties on the trigger efficiency quoted above.

\subsection{Effect of higher-order QCD corrections}
\label{sec:NLOcorr}
For the $\Higgsdecay$ sample with $m_{\Higgs} = 125$\GeVcc, the leptons from the \X boson decay have a
combined efficiency of only a few percent for passing the lepton \pt requirements. For this reason the signal
efficiency at this mass is sensitive to the modelling of the Higgs boson \pt spectrum, which may in turn be
influenced by higher-order QCD corrections. To evaluate this effect, we reweight the LO Higgs boson \pt spectrum from
our signal sample to match the corresponding Higgs boson \pt spectrum evaluated at NLO~\cite{Nason:2004rx,Frixione:2007vw,Alioli:2010xd}. For $m_{\Higgs} =
125$\GeVcc and $m_{X} = 20\,(50)$\GeVcc the signal efficiency changes by 5\%,(7\%). This change is taken as
an additional systematic uncertainty in the efficiency for the case $m_{\Higgs} = 125$\GeVcc. For the larger
\Higgs masses that we consider, and also for the neutralino channel, where a similar study was performed,
the corresponding systematic uncertainty is below 0.5\%, and hence neglected.

\section{Results}
\label{sec:results}

Events from background sources are equally likely to populate the signal and control regions, whereas any
events arising from LL particles will populate almost exclusively the signal region. In consequence, the
presence  of a signal in the data would reveal itself as a statistically significant excess of events in the
signal region compared to the control region. After all selection requirements are applied, no events are found
in the signal or control regions in either the electron or muon channel. There is thus no statistically
significant excess. The \dxysigma distributions of events in the signal and control regions were shown in
Fig.~\ref{fig:D0SigmaDataMC}.

We set 95\% confidence level (\CL) upper limits on the signal processes using the Bayesian method described in
Ref.~\cite{LHC:HiggsCombination}. The limits are determined from a comparison of the number of events observed in the signal region with the number
expected in the signal plus background hypothesis.

The limit calculation takes into account the systematic uncertainties in the signal yield, described in
Section~\ref{sec:Systematics}, by introducing nuisance parameters for each of the uncertainties that are marginalized through an integration over their
log-normal prior distributions. The expected number of background events $\mu_B$ in the control region, and
hence also in the signal region, is an additional nuisance parameter. It is constrained by the observed number
of events $N_C$ in the control region. Its probability distribution $p(\mu_B | N_C)$ is given by:
\begin{linenomath}
\begin{equation*}
p(\mu_B | N_C) = \frac{\mu_B^{N_C}}{N_C !}\exp(-\mu_B),
\end{equation*}
\end{linenomath}
as can be shown using Bayesian methodology assuming a flat prior in
$\mu_B$~\cite{LHC:HiggsCombination}. The expected background in the signal region may differ from that in the control region,
as a result of tracker misalignment. This is taken into account as described in Section~\ref{sec:bkgndest}, by including an
appropriate systematic uncertainty, and by evaluating the limits twice, once with and once without correcting the track
impact parameters for tracker misalignment, and taking the worse of these two sets of limits as the result.

If a genuine signal were present, it would give rise to an excess of events in the signal region
with an expected number of:
\begin{linenomath}
\begin{equation}\label{eqnlimitA}
\mu_S    =     \Lumi\sigma\left[ 2 \BR (1-\BR) \epsilon_{1} + \epsilon_{2} \BR^2 \right](1-f),
\end{equation}
\end{linenomath}
where $\Lumi$ is the integrated luminosity, $\epsilon_{(1,2)}$ are the signal efficiencies defined in Section~\ref{sec:RecoAndSelection},
$\sigma$ is the production cross section of \Higgsdecay (or \sqasq) and \BR is the
branching fraction for the decay \Xdecay (or \sqdecay). The parameter $f$ is the mean number of signal events
expected to fall in the control region for each signal event in the signal region. This fraction is very small, being
less than 3\% for all the $\Xdecay$ samples and less than 5\% for all
the $\chidecay$ samples considered here. Its effect is to reduce slightly the effective signal
efficiency, by causing some of the signal to be misinterpreted as background.
One expects $\epsilon_{2}\ge 1-(1-\epsilon_{1})^2$, where the two terms are equal if the efficiency to
select each of the two LL particles in an event is independent of the other, or the first term is larger
if the presence of one LL particle increases the efficiency to select the other (as can happen if one lepton
from each causes the event to trigger). Assuming $\epsilon_{2} = 1-(1-\epsilon_{1})^2$, which is conservative
since it minimizes the value of $\mu_S$, transforms Eq.~(\ref{eqnlimitA}) into:
\begin{linenomath}
\begin{equation}\label{eqnlimitB}
\mu_S   =   2 \Lumi\sigma\BR\epsilon_{1}\left[1 - \frac{1}{2}\BR\epsilon_{1}\right](1-f)\ .
\end{equation}
\end{linenomath}
Since $\mu_S$ in Eq.~(\ref{eqnlimitB}) depends not only on $\sigma\BR$, but also on \BR, the upper limits on
$\sigma\BR$ depend on the assumed value of \BR, scaling approximately as the expression $1/[1 - \frac{1}{2}\BR\epsilon_{1}]$.
The upper limits are thus best for low values of \BR, though the dependence of the limits on \BR is weak, particularly
if $\epsilon_{1}$ is small. We set the value of \BR equal to unity in the expression in square brackets, so as to obtain conservative limits that are valid
for any value of \BR.

\begin{figure*}[hbtp]
\centering
\includegraphics[width=0.49\textwidth]{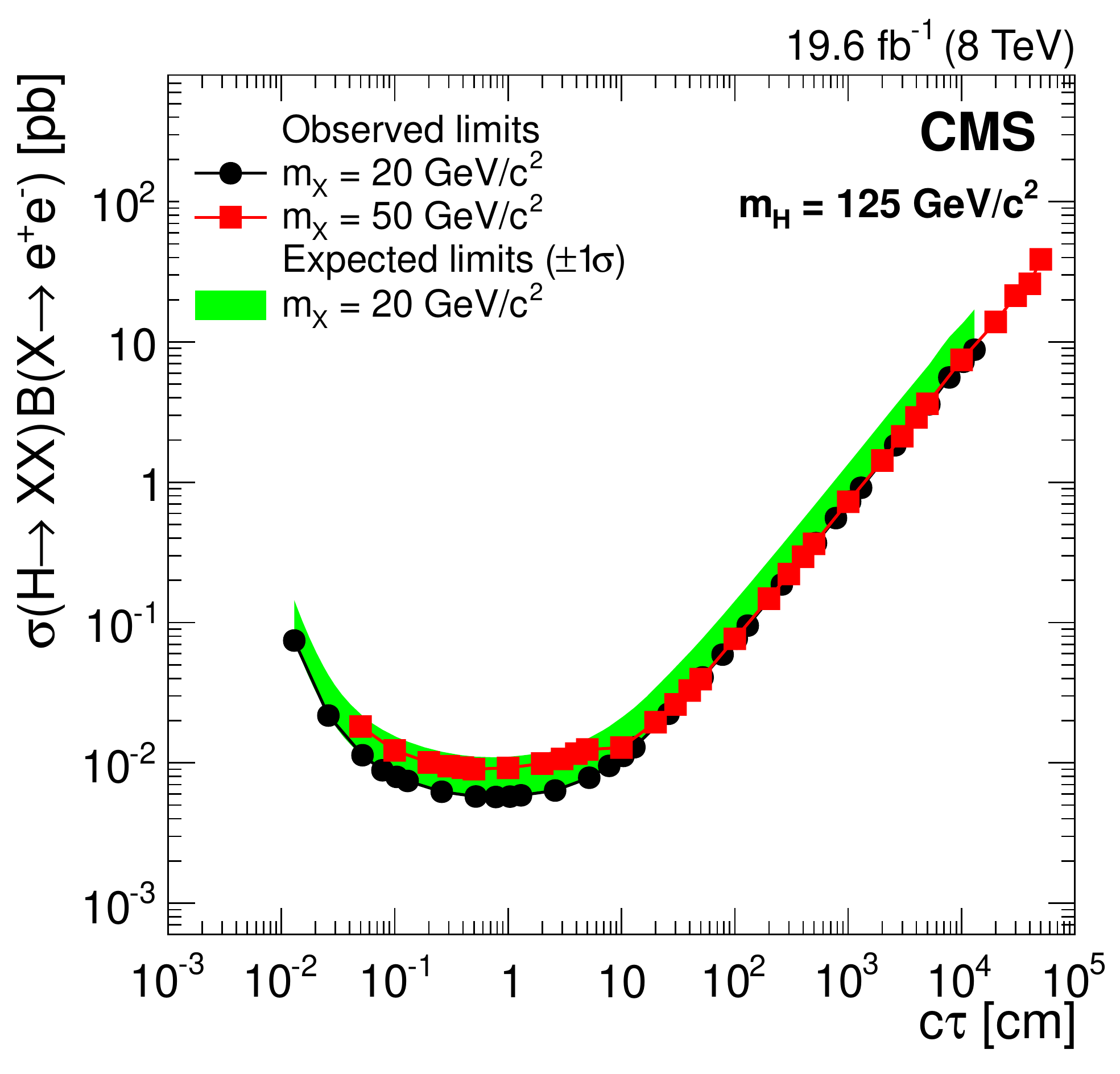}
\includegraphics[width=0.49\textwidth]{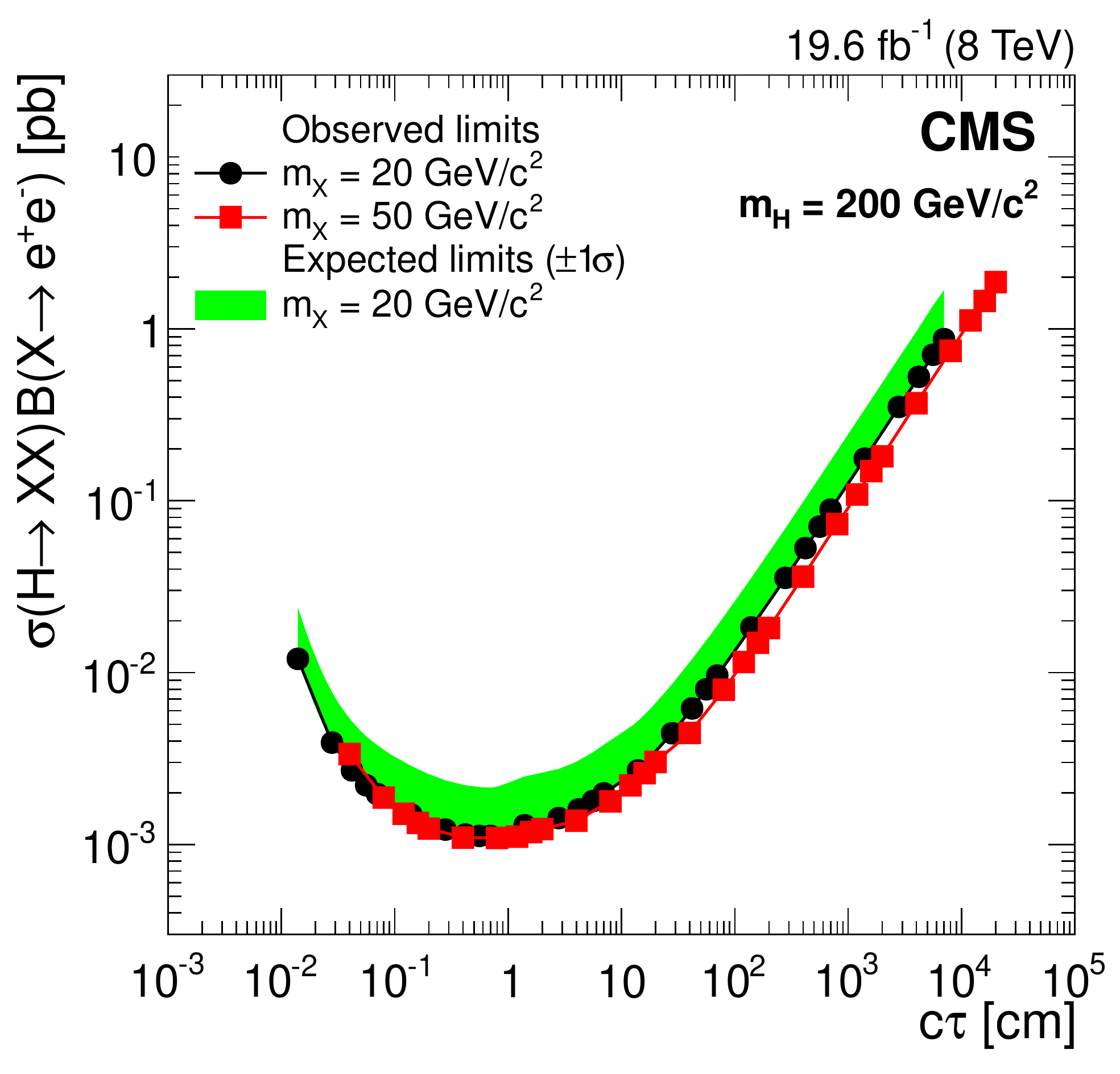}
\includegraphics[width=0.49\textwidth]{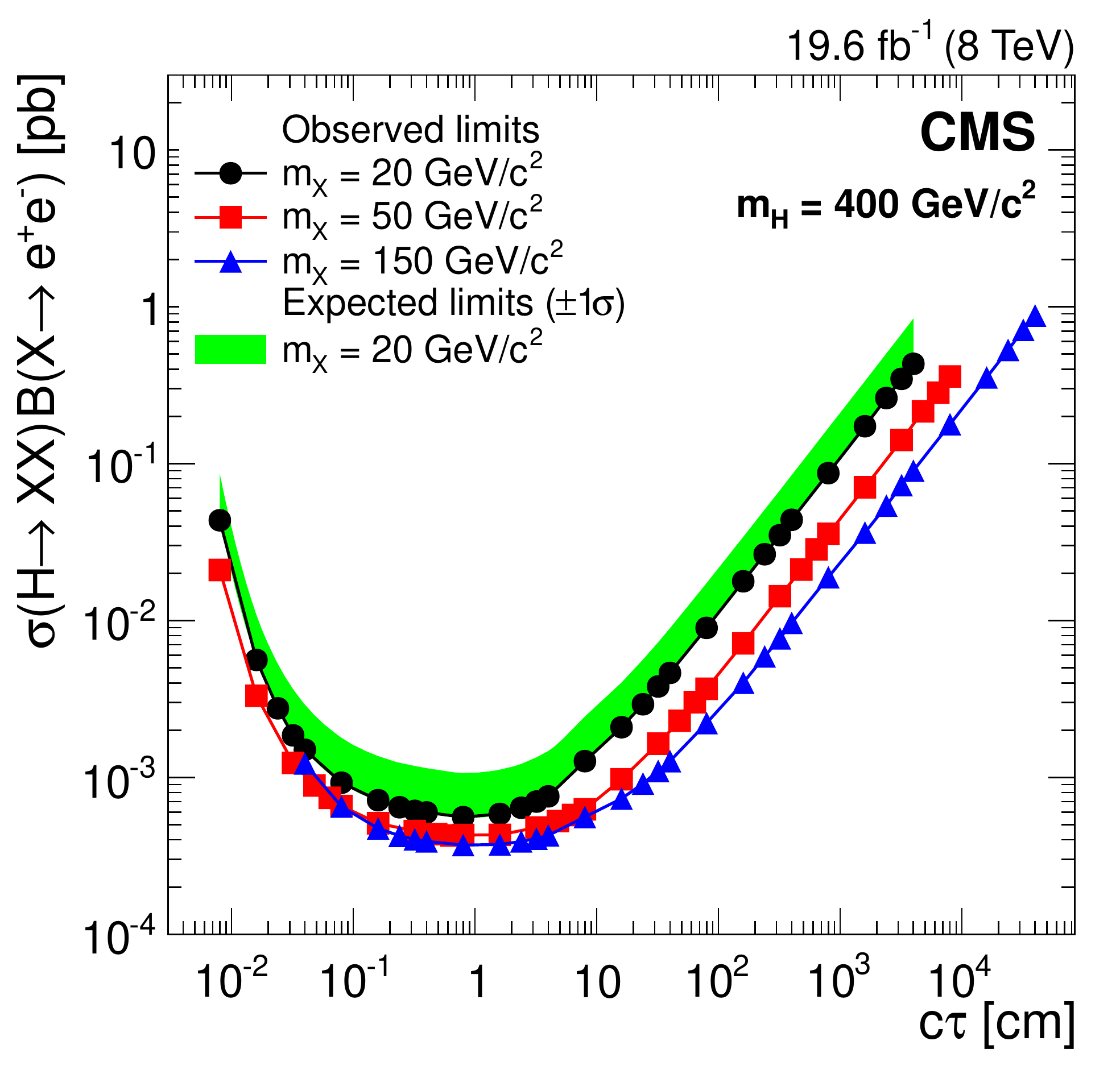}
\includegraphics[width=0.49\textwidth]{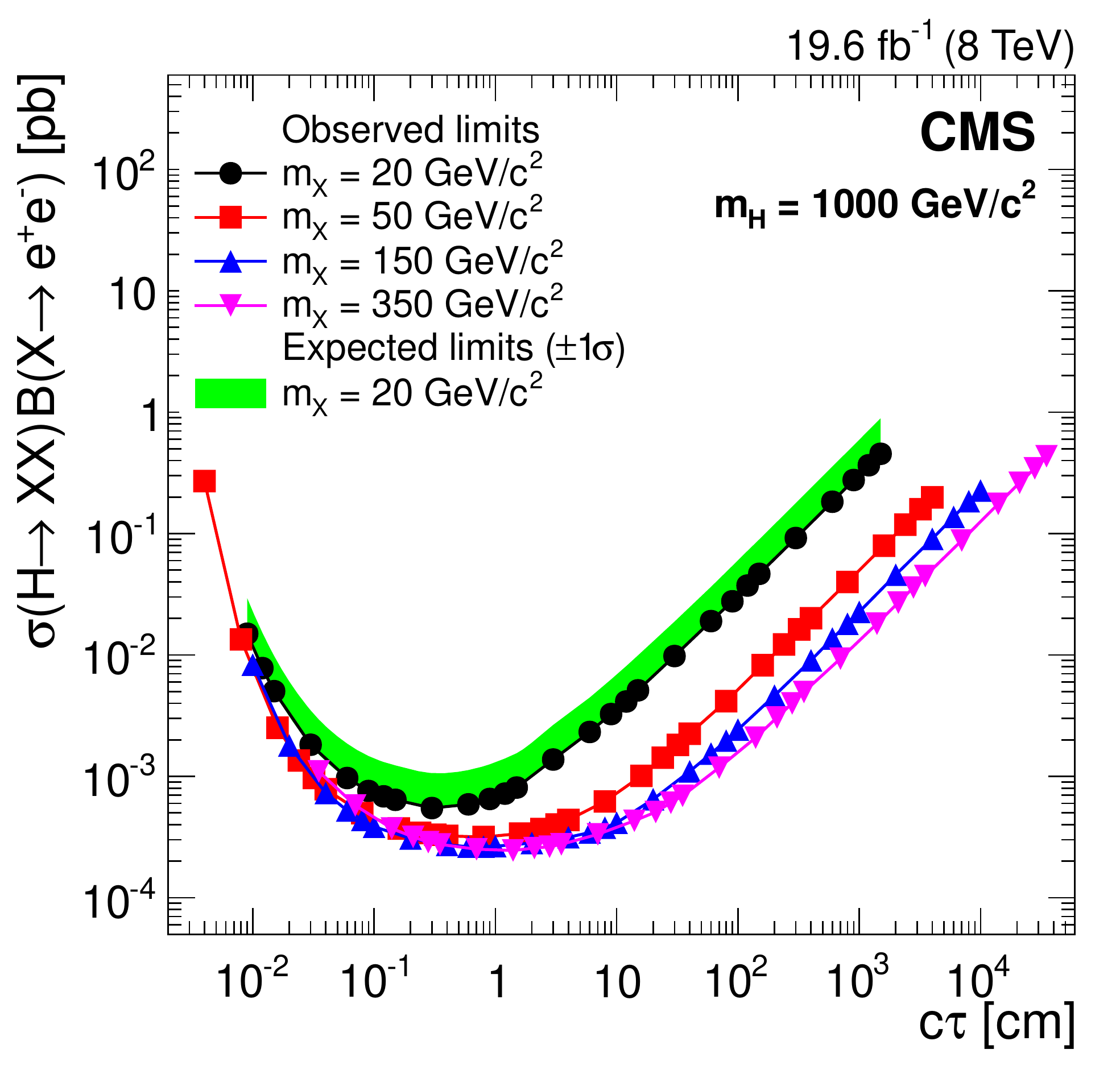}
\caption{The 95\% \CL upper limits on $\sigma(\Higgsdecay)\BR(\Xdecayee)$, as a function of the mean proper decay length of the \X boson,
for Higgs boson masses of 125\GeVcc (top left),
200\GeVcc (top right), 400\GeVcc (bottom left), and 1000\GeVcc (bottom right). In each plot, results are shown for several \X boson
mass hypotheses. The shaded band shows the ${\pm}1\sigma$ range of variation of the expected 95\% \CL limits for the case of a 20\GeVcc \X boson mass.
Corresponding bands for the other \X boson masses, omitted for clarity of presentation, show similar agreement with the respective observed limits.}
\label{fig:FinLimitsCountEE}
\end{figure*}

\begin{figure*}[hbtp]
\centering
\includegraphics[width=0.49\textwidth]{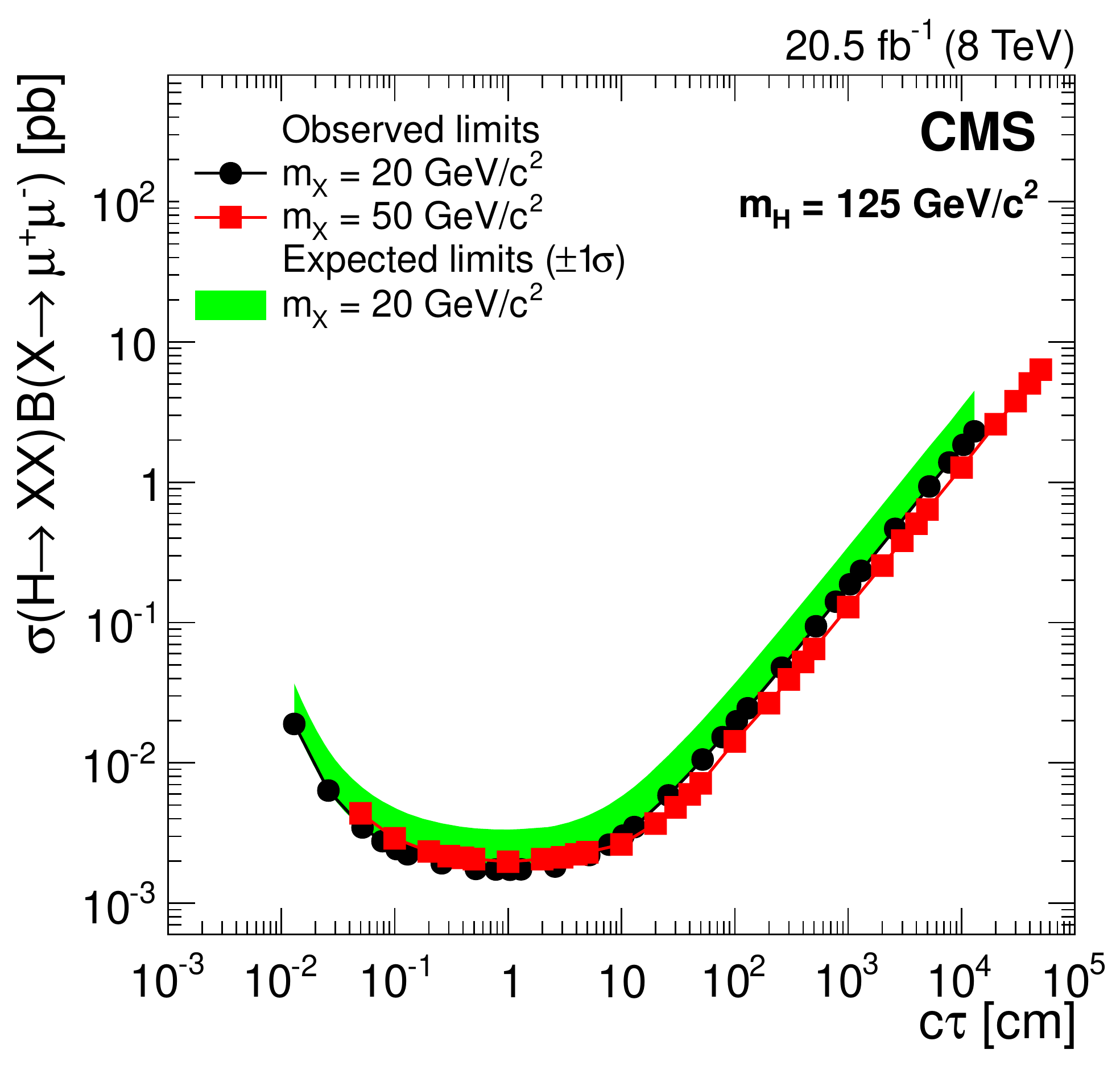}
\includegraphics[width=0.49\textwidth]{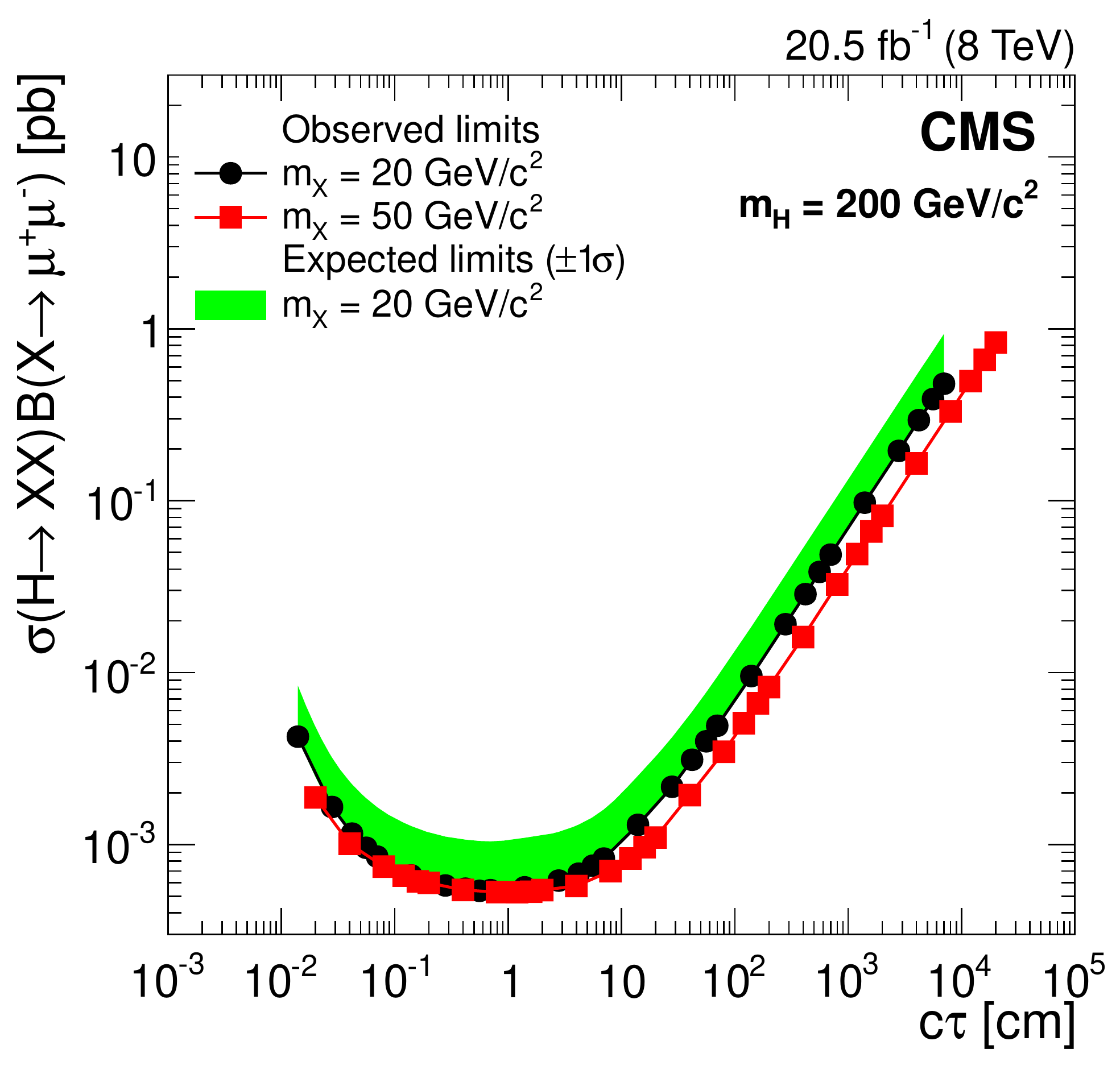}
\includegraphics[width=0.49\textwidth]{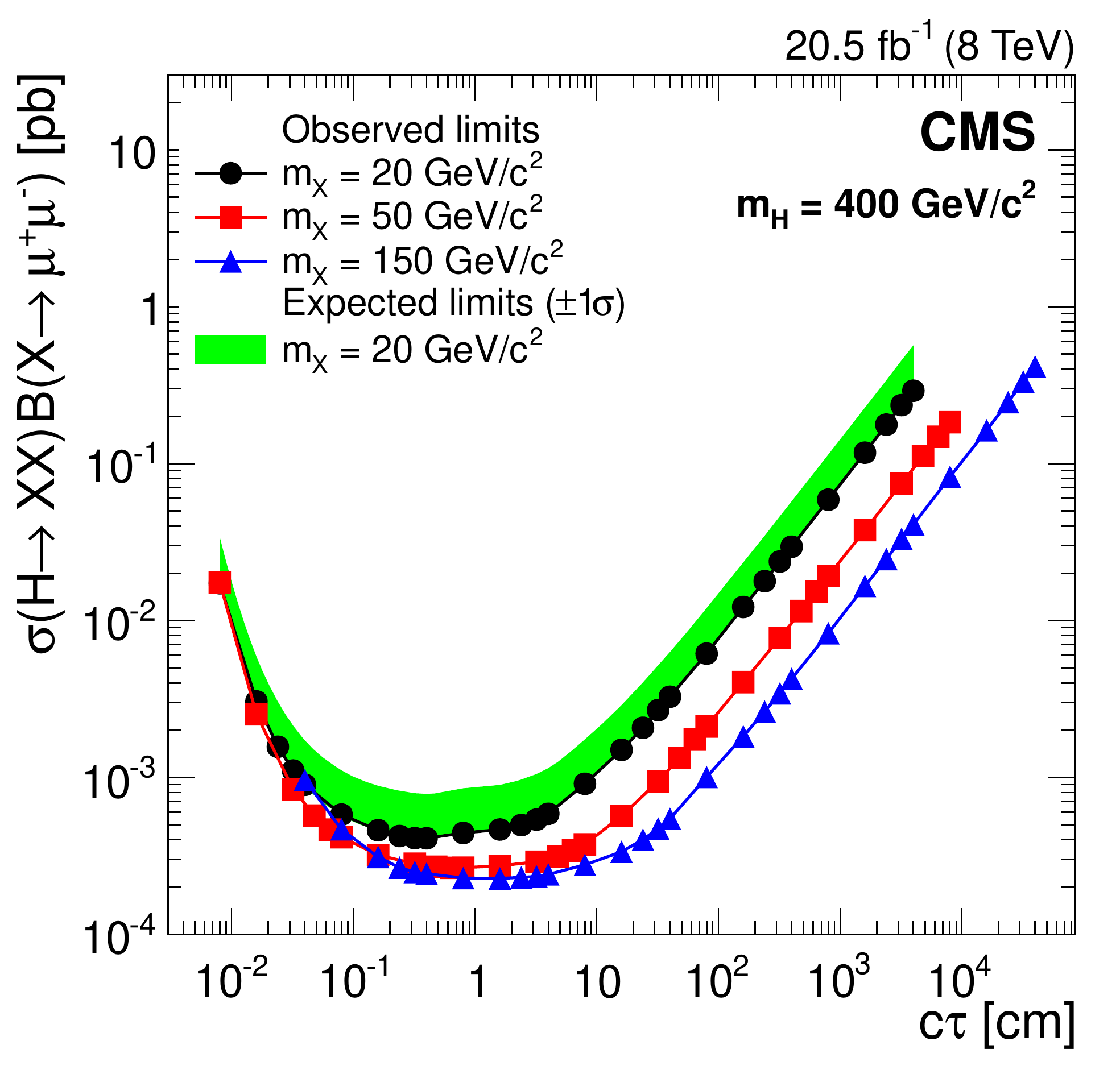}
\includegraphics[width=0.49\textwidth]{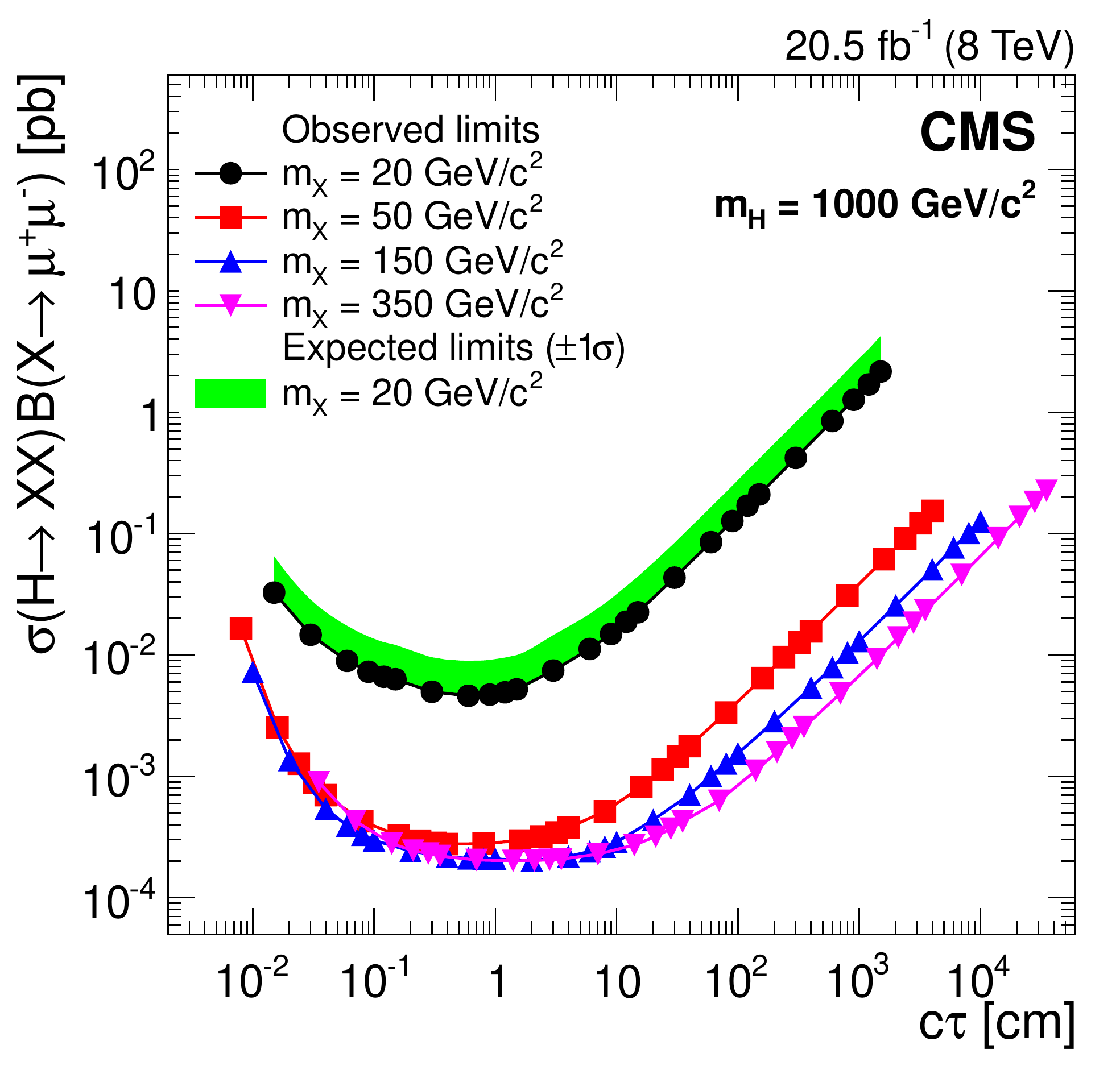}
\caption{The 95\% \CL upper limits on $\sigma(\Higgsdecay)\BR(\Xdecaymumu)$, as a function of the mean proper decay length of the \X boson,
for Higgs boson masses of 125\GeVcc (top left),
200\GeVcc (top right), 400\GeVcc (bottom left), and 1000\GeVcc (bottom right). In each plot, results are shown for several \X boson
mass hypotheses. The shaded band shows the ${\pm}1\sigma$ range of variation of the expected 95\% \CL limits for the case of a 20\GeVcc \X boson mass.
Corresponding bands for the other \X boson masses, omitted for clarity of presentation, show similar agreement with the respective observed limits.}
\label{fig:FinLimitsCountMuMu}
\end{figure*}

\begin{figure*}[hbtp]
\centering
\includegraphics[width=0.49\textwidth]{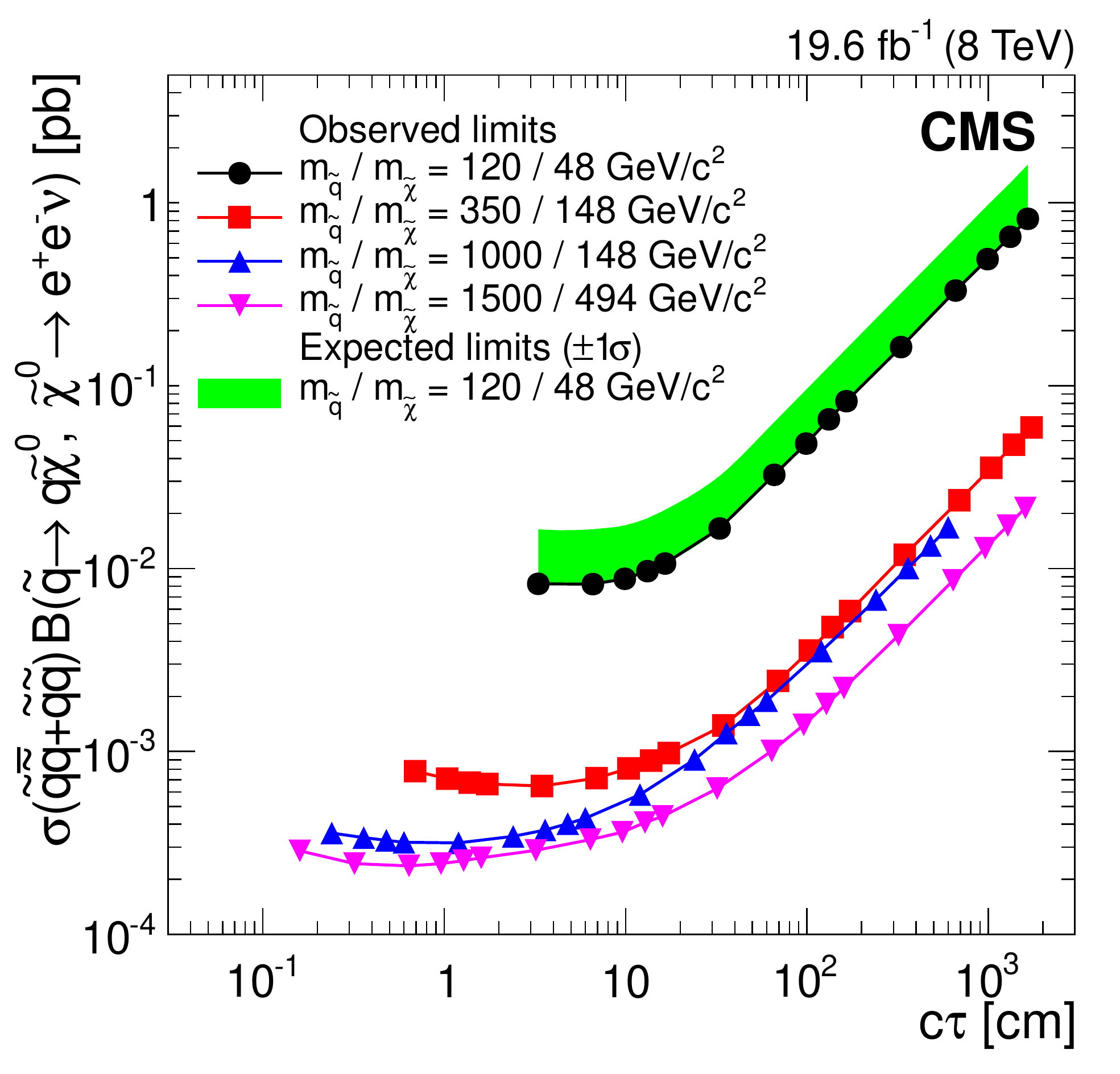}
\includegraphics[width=0.49\textwidth]{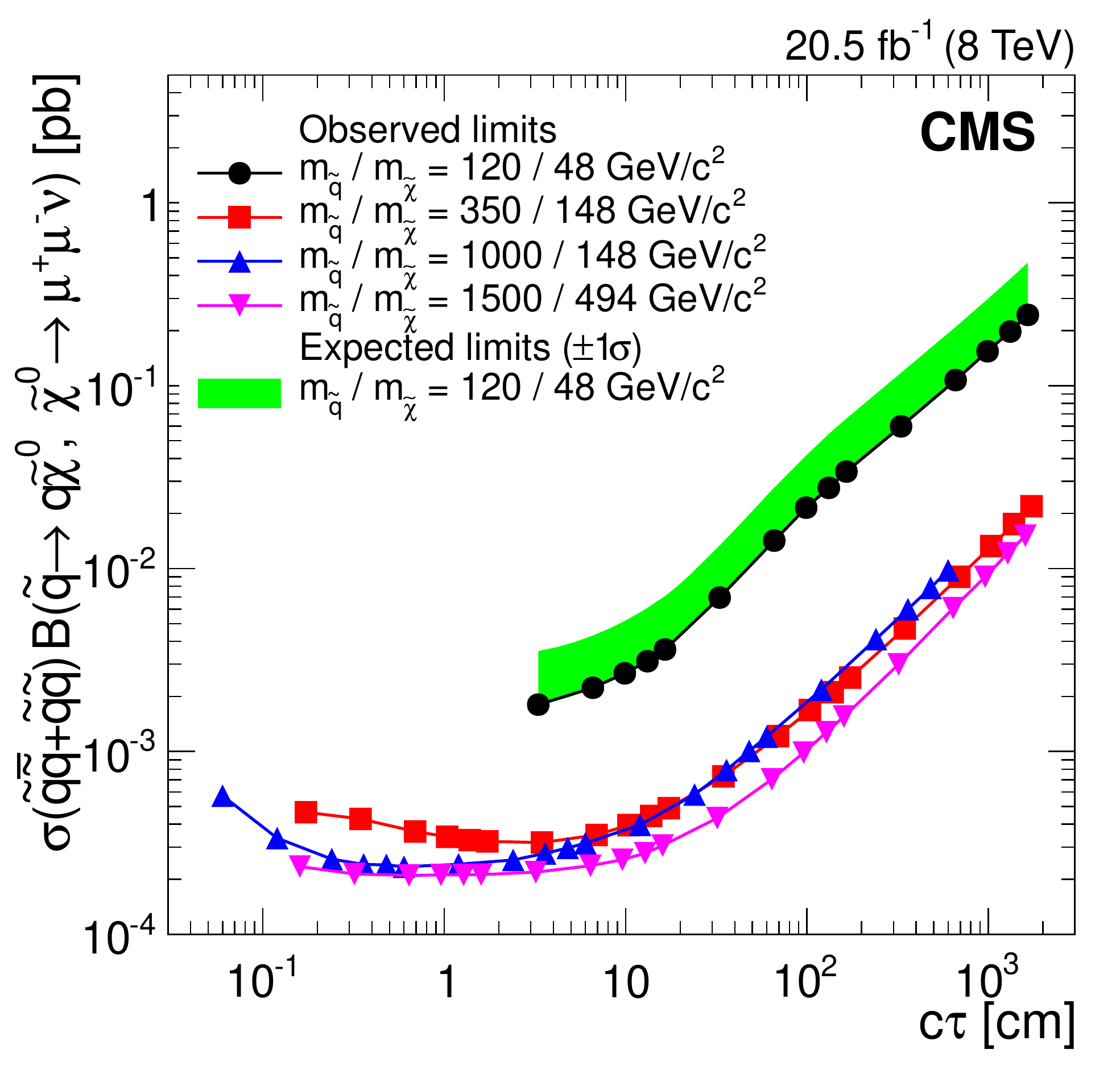}
\caption{The 95\% \CL upper limits on $\sigma(\sqasq)\BR(\sqdecay)$ for the electron (left), and muon (right) channels, as a function of the mean proper decay length of the neutralino. The shaded band shows the ${\pm}1\sigma$ range of variation of the expected 95\% \CL limits for the case of a 120\GeVcc squark and a 48\GeVcc neutralino mass.  Corresponding bands for the other squark and neutralino masses, omitted for clarity of presentation, show similar agreement with the respective observed limits.}
\label{fig:FinLimitsCountNeutralino}
\end{figure*}

For each combination of the \Higgs and \X boson masses that are modelled, and for a range of mean proper decay lengths $\ctau$ of the \X boson, 95\% \CL upper limits on $\sigma(\Higgsdecay)\BR(\Xdecay)$ are calculated.
The observed limits for the electron and muon channels are shown in Figs.~\ref{fig:FinLimitsCountEE} and
\ref{fig:FinLimitsCountMuMu}, respectively. The less stringent limits for the muon
channel in the $m_{\Higgs} = 1000\GeVcc$, $m_{\X} = 20\GeVcc$ case are caused by low trigger efficiency for nearby muons,
and the consequent $\Delta R$ requirement. The corresponding limits on
$\sigma(\sqasq)\BR(\sqdecay)$ are shown in Fig.~\ref{fig:FinLimitsCountNeutralino}.
The shaded band in each of these plots shows the ${\pm}1\sigma$ range of variation of the
expected 95\% \CL limits, illustrated for one choice of masses. All the observed limits are consistent with
the corresponding expected ones.

At $\sqrt{s}=8$\TeV, the theoretical cross sections for SM Higgs boson production through the dominant
gluon-gluon fusion mechanism are 19.3, 7.1, 2.9, and 0.03\unit{pb} for Higgs boson masses of 125, 200, 400,
and 1000\GeVcc, respectively~\cite{Heinemeyer:2013tqa}. The theoretical cross sections for \sqasq
production are 2590, 10, 0.014, and 0.00067\unit{pb} for \squark masses of 120, 350, 1000, and 1500\GeVcc, as
evaluated with the \PROSPINO generator~\cite{Beenakker:1996ch} assuming a gluino mass of 5\TeVcc.
The observed limits on $\sigma\BR$ are usually well below these theoretical cross sections, implying
that nontrivial bounds are being placed on the decay modes involving LL particles, probing, for example,
branching fractions as low as $10^{-4}$ and $10^{-6}$ in the Higgs and supersymmetric models, respectively.

\begin{figure*}[hbtp]
\centering
\includegraphics[width=0.49\textwidth]{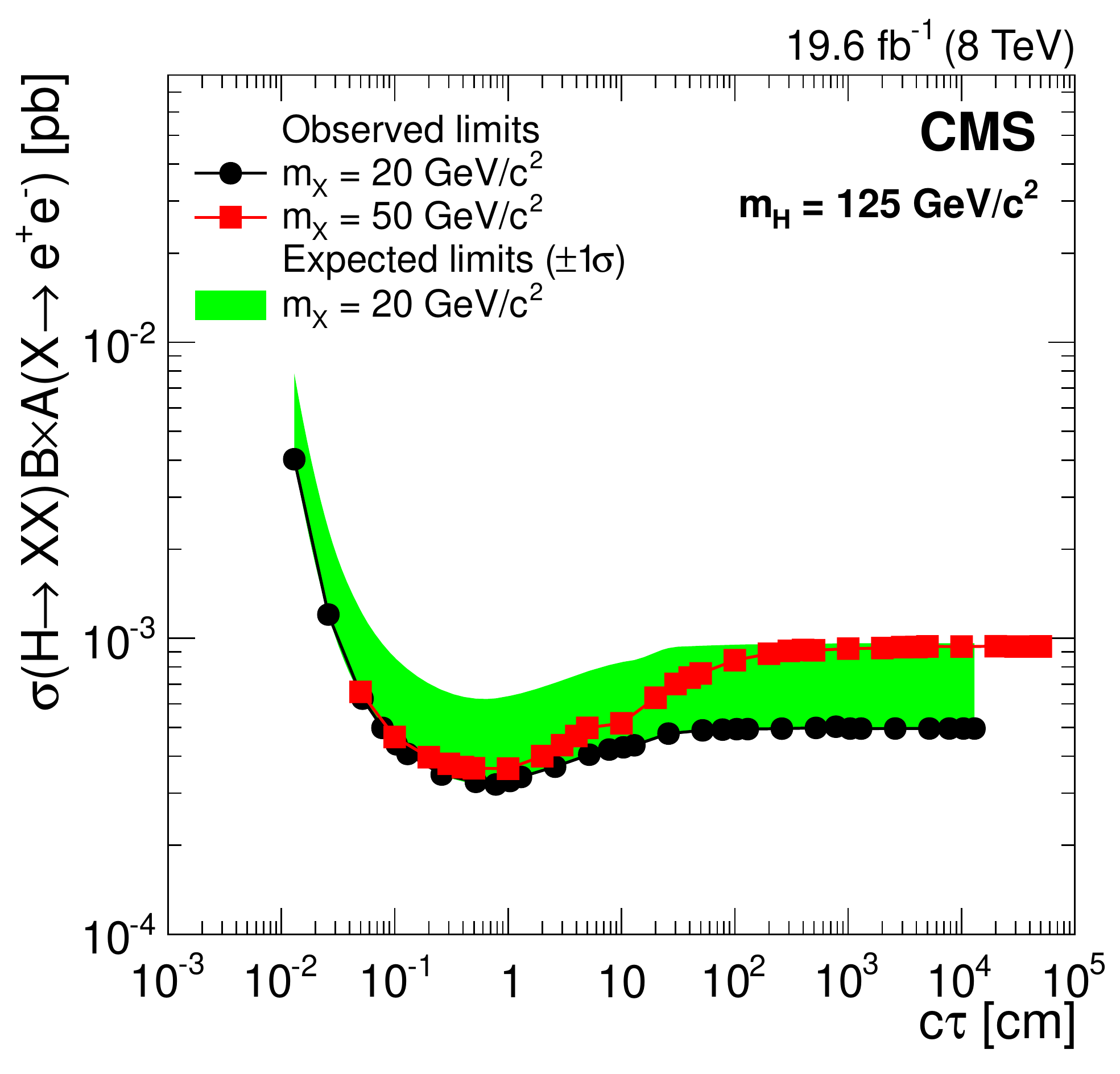}
\includegraphics[width=0.49\textwidth]{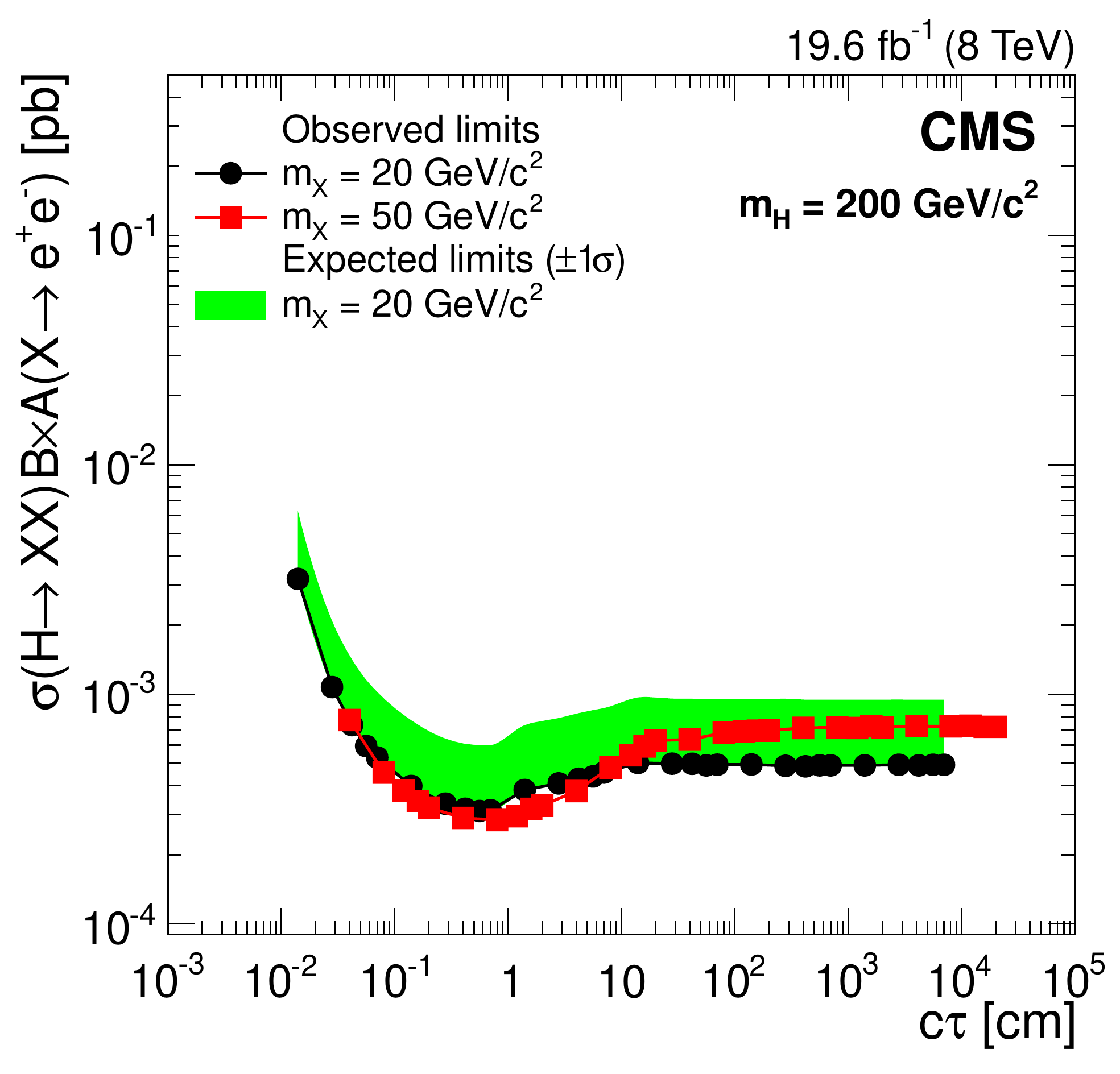}
\includegraphics[width=0.49\textwidth]{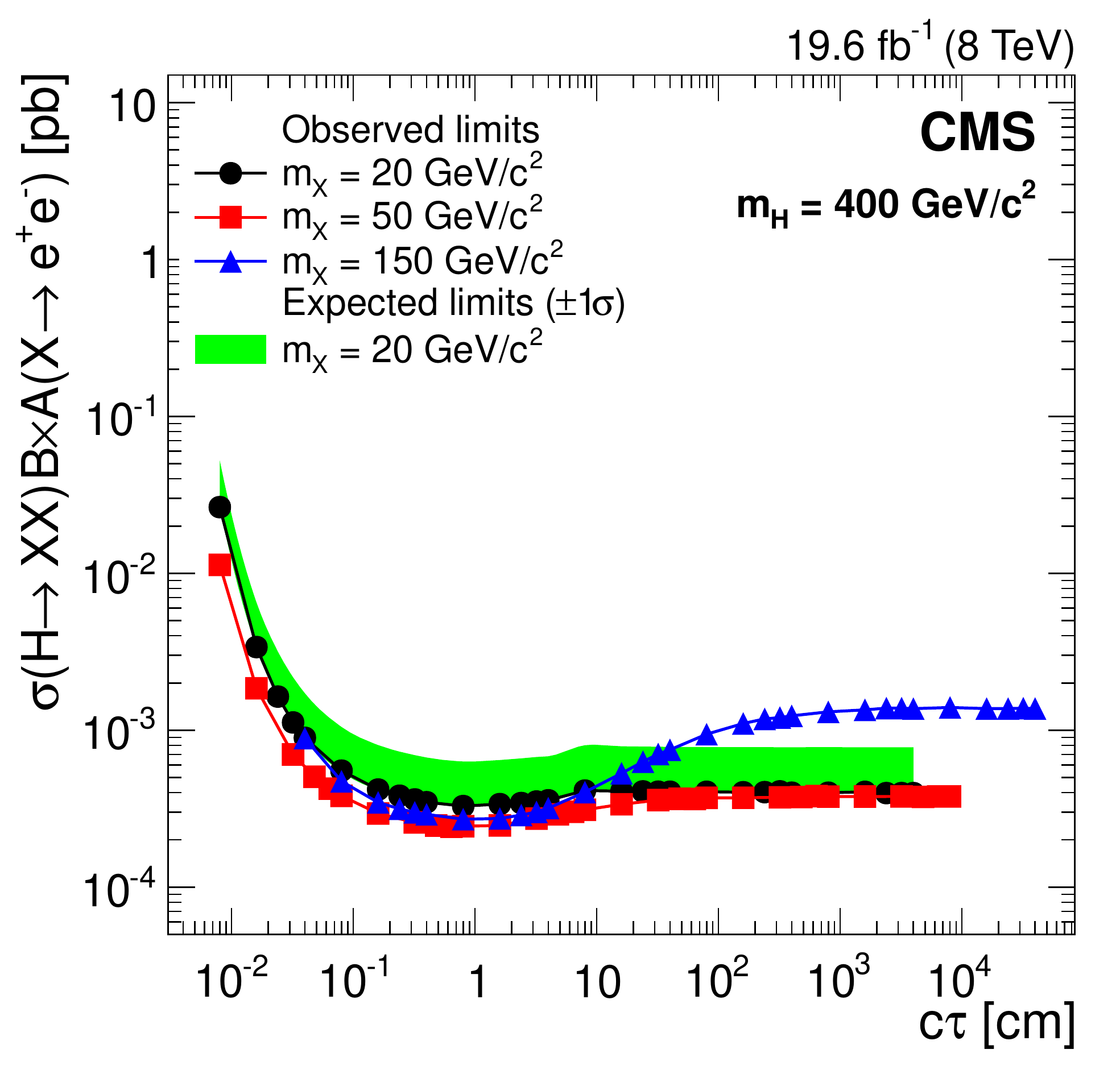}
\includegraphics[width=0.49\textwidth]{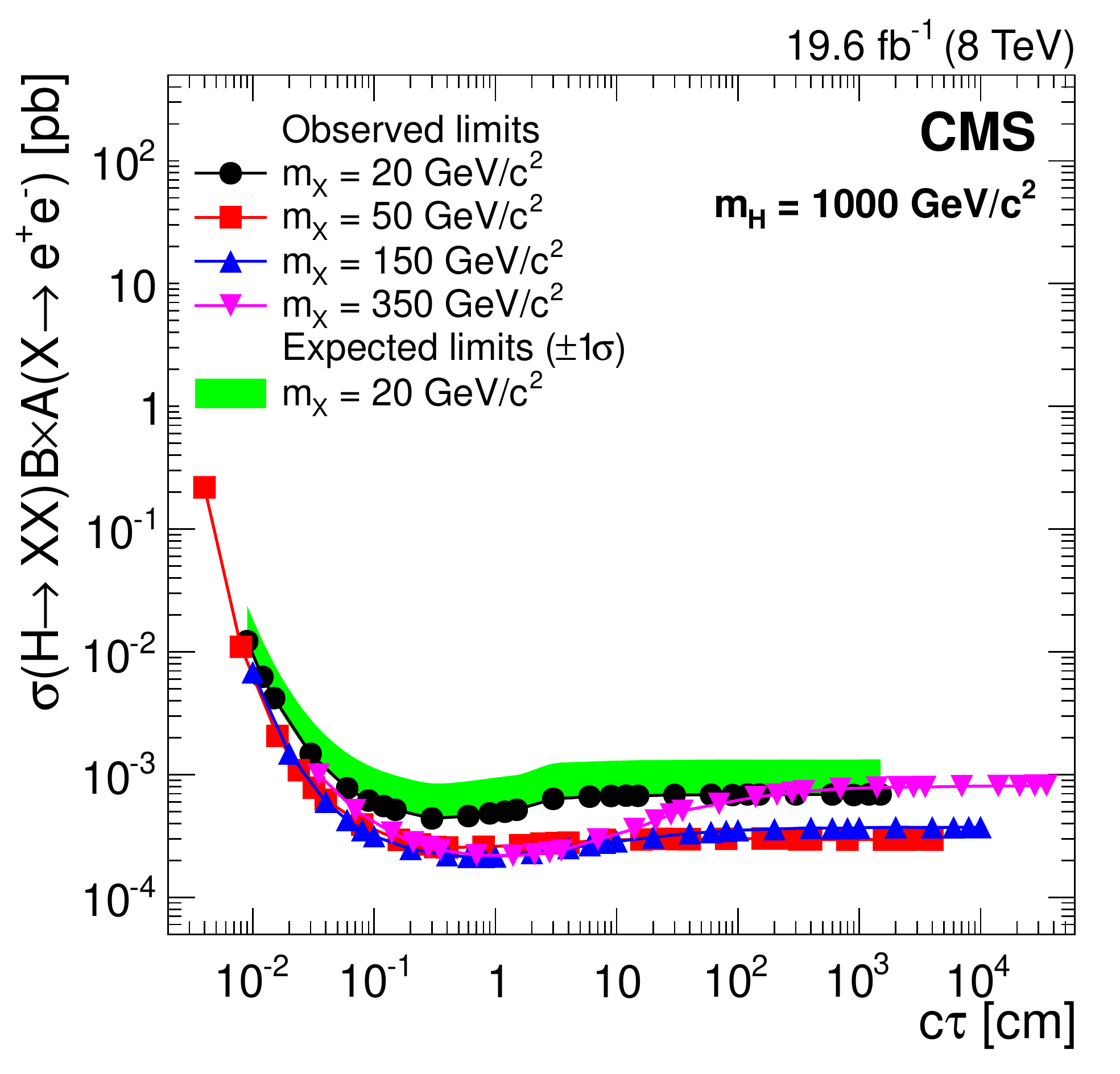}
\caption{The 95\% \CL upper limits on $\sigma(\Higgsdecay)\BR(\Xdecayee)A(\Xdecayee)$, as a function of the mean proper decay length of the \X boson,
for Higgs boson masses of 125\GeVcc (top left),
200\GeVcc (top right), 400\GeVcc (bottom left), and 1000\GeVcc (bottom right). In each plot, results are shown for several \X boson
mass hypotheses. The shaded band shows the ${\pm}1\sigma$ range of variation of the expected 95\% \CL limits for the case of a 20\GeVcc \X boson mass.
Corresponding bands for the other \X boson masses, omitted for clarity of presentation, show similar agreement with the respective observed limits.}
\label{fig:FinLimitsCountAccEE}
\end{figure*}

\begin{figure*}[hbtp]
\centering
\includegraphics[width=0.49\textwidth]{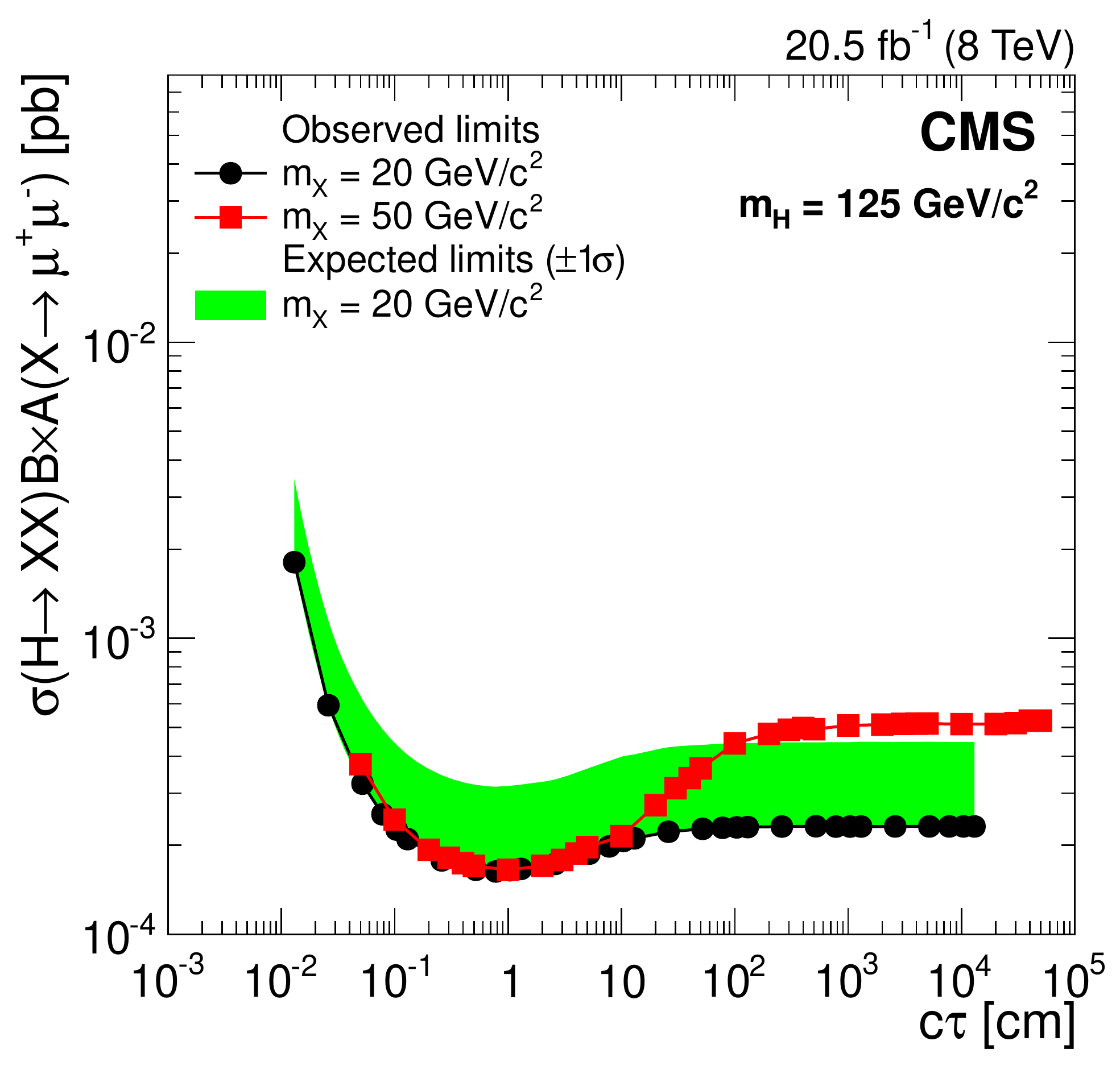}
\includegraphics[width=0.49\textwidth]{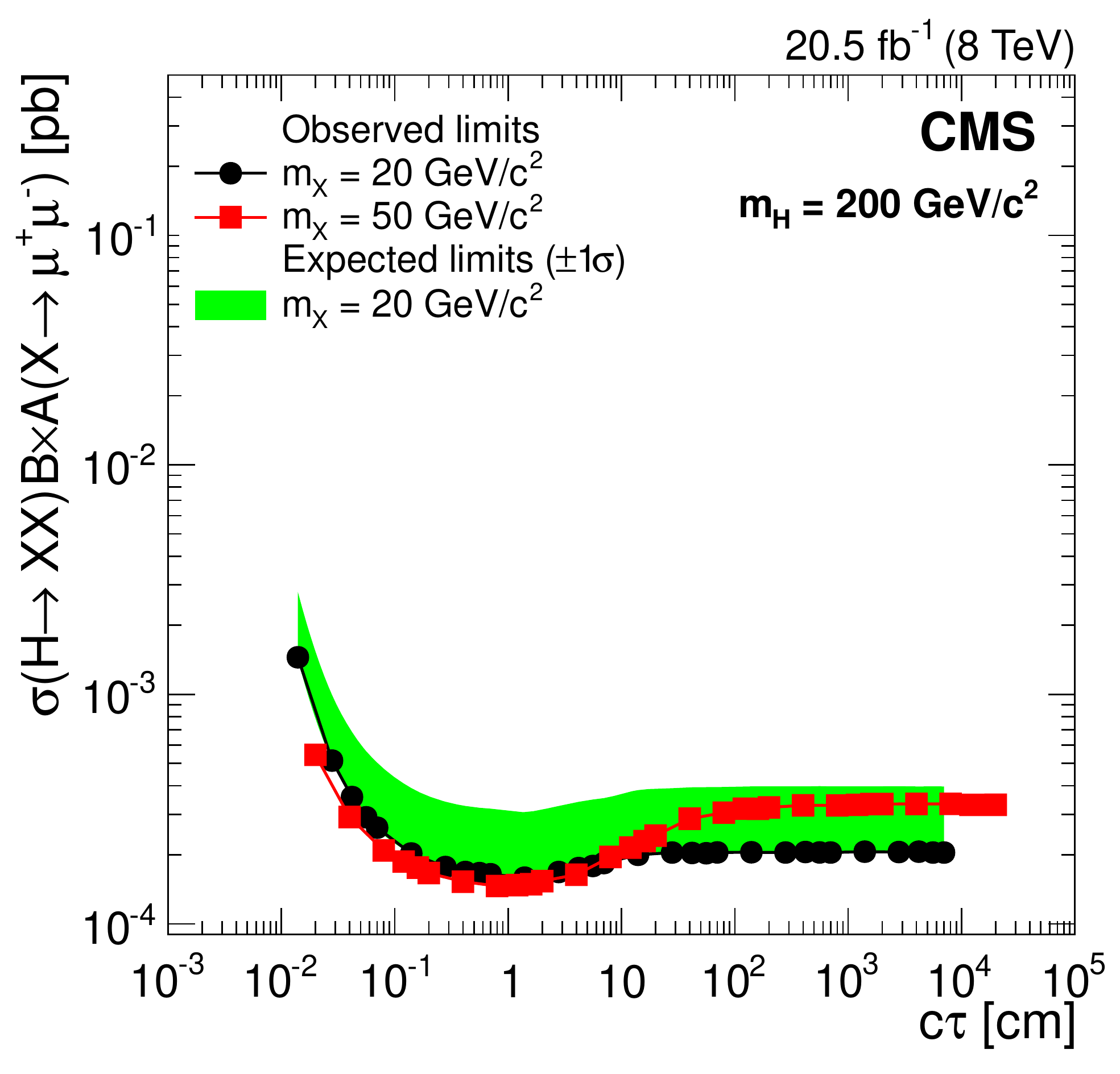}
\includegraphics[width=0.49\textwidth]{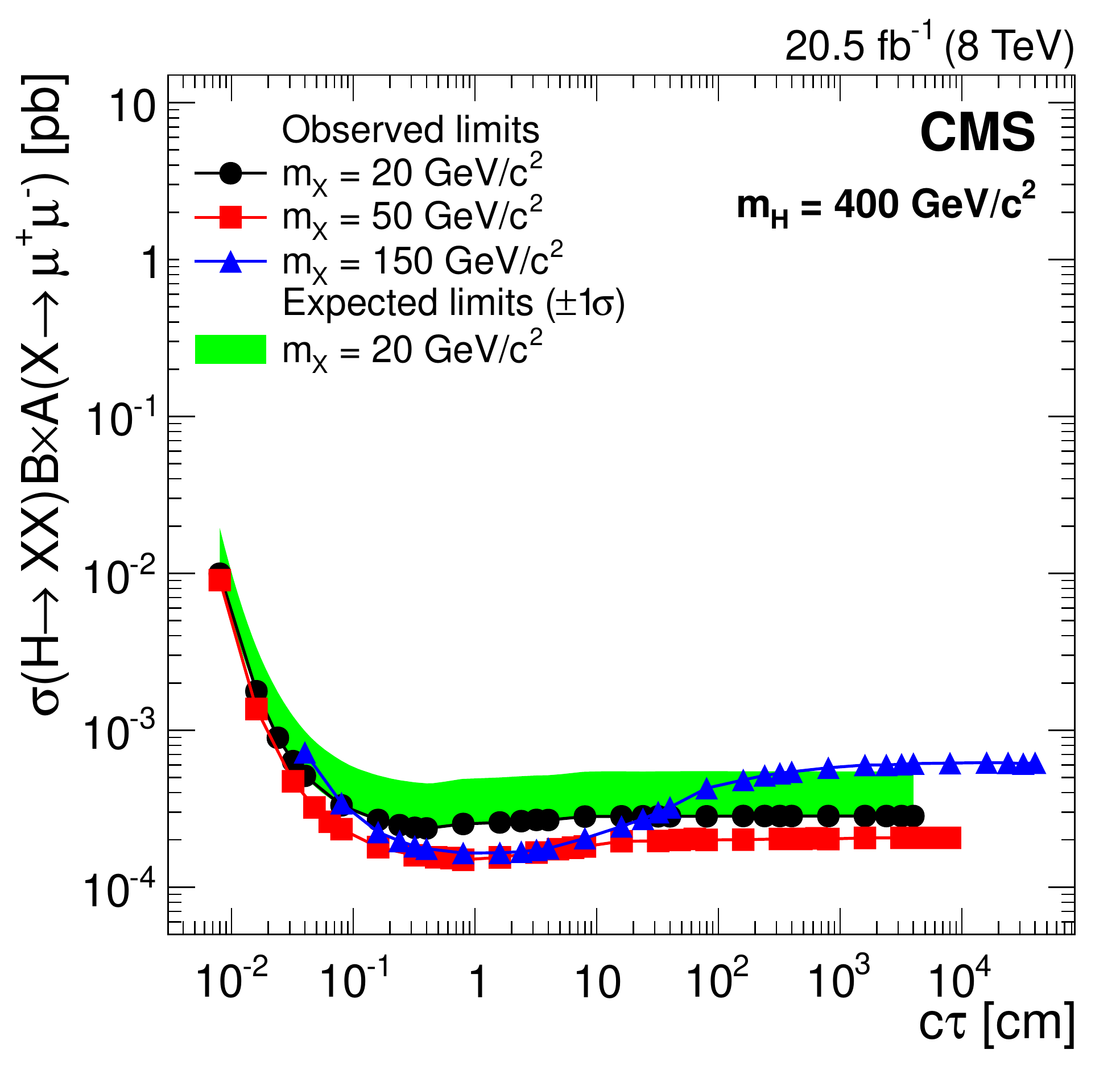}
\includegraphics[width=0.49\textwidth]{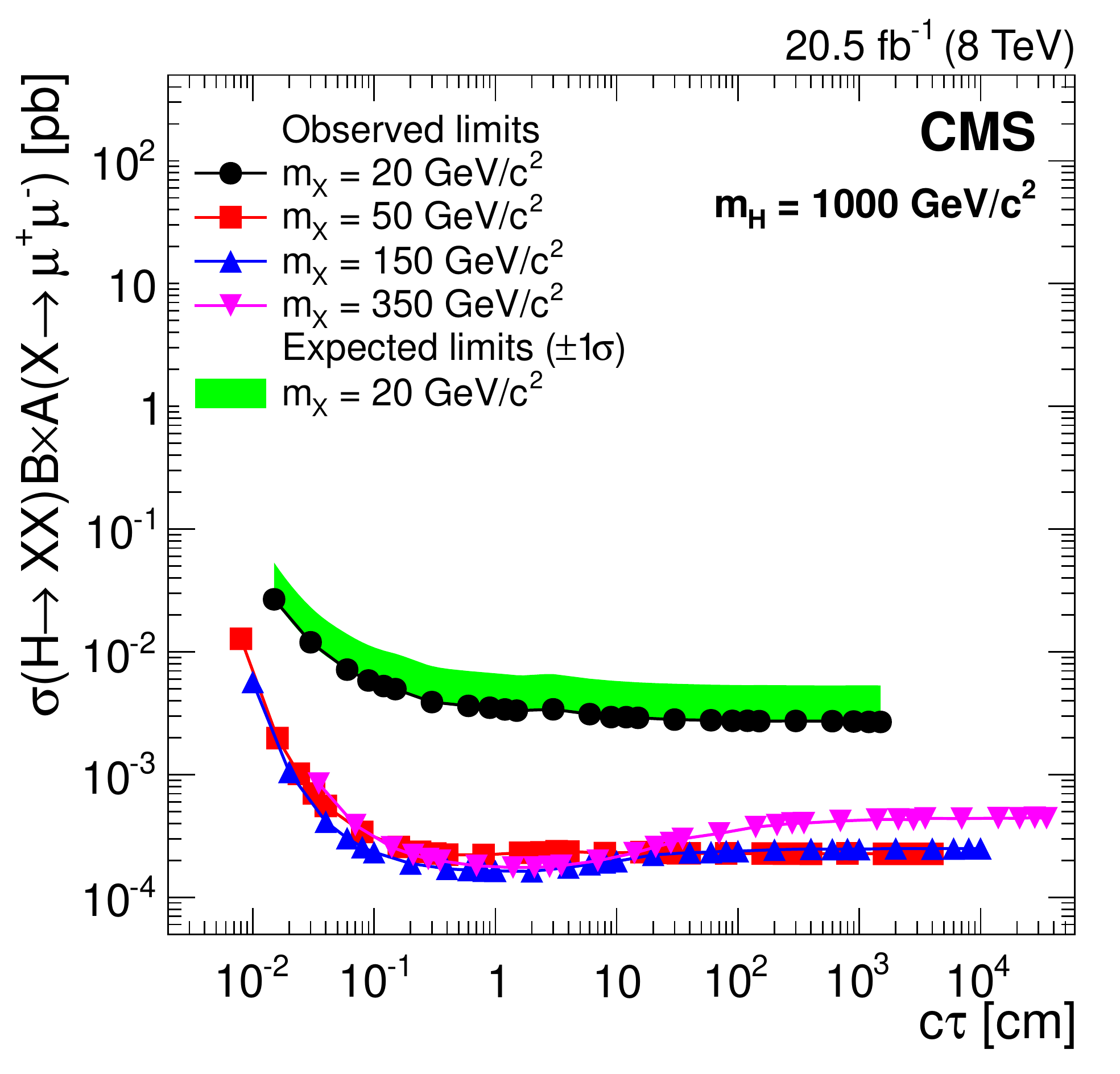}
\caption{The 95\% \CL upper limits on $\sigma(\Higgsdecay)\BR(\Xdecaymumu)A(\Xdecaymumu)$, as a function of the mean proper decay length of the \X boson,
for Higgs boson masses of 125\GeVcc (top left),
200\GeVcc (top right), 400\GeVcc (bottom left), and 1000\GeVcc (bottom right). In each plot, results are shown for several \X boson
mass hypotheses. The shaded band shows the ${\pm}1\sigma$ range of variation of the expected 95\% \CL limits for the case of a 20\GeVcc \X boson mass.
Corresponding bands for the other \X boson masses, omitted for clarity of presentation, show similar agreement with the respective observed limits.}
\label{fig:FinLimitsCountAccMuMu}
\end{figure*}

\begin{figure*}[hbtp]
\centering
\includegraphics[width=0.49\textwidth]{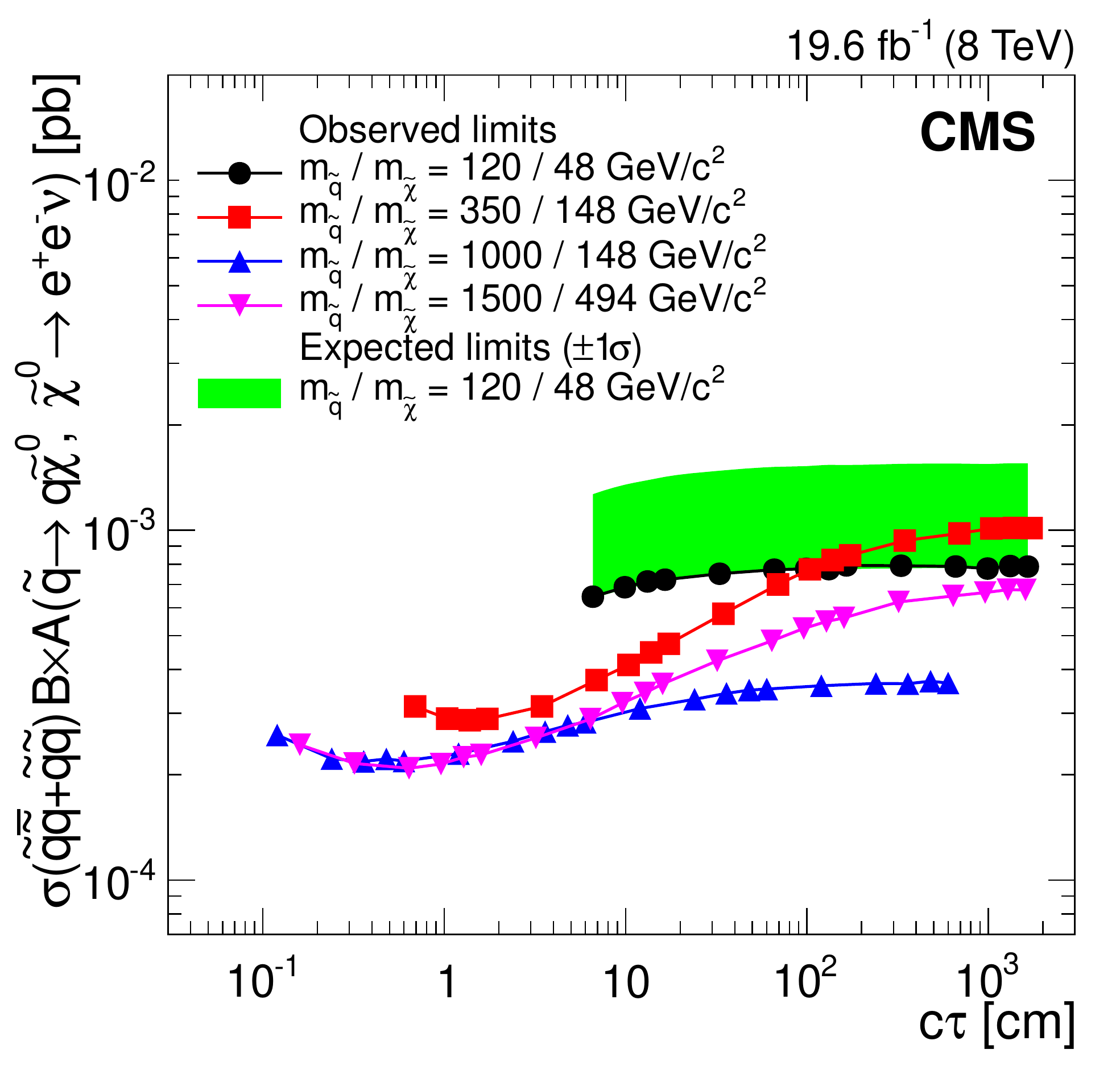}
\includegraphics[width=0.49\textwidth]{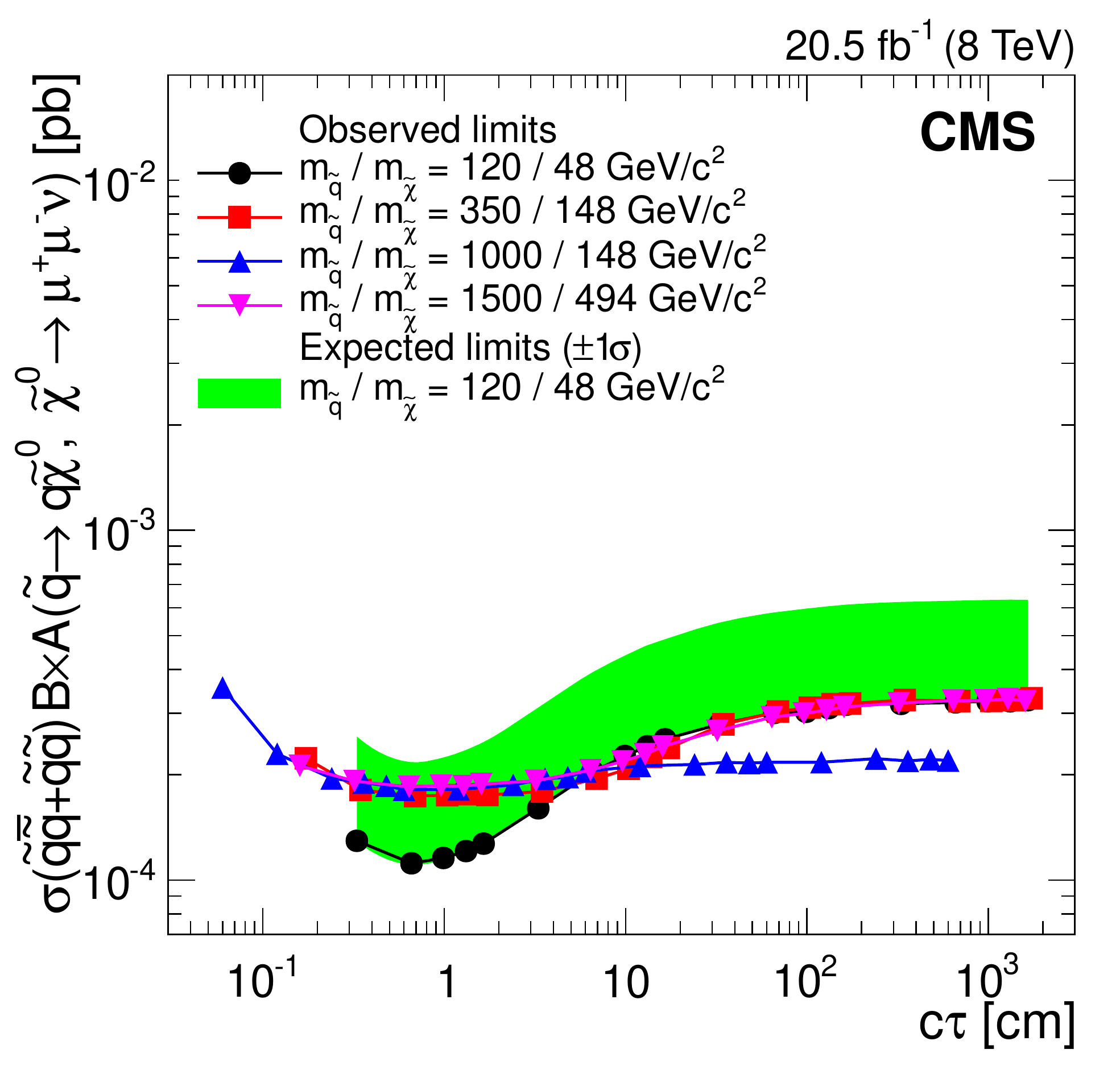}
\caption{The 95\% \CL upper limits on $\sigma(\sqasq)\BR(\sqdecay)A(\sqdecay)$ for the electron (left), and muon (right) channels, as a function of the mean proper decay length of the neutralino. The shaded band shows the ${\pm}1\sigma$ range of variation of the expected 95\% \CL limits for the case of a 120\GeVcc squark and a 48\GeVcc neutralino mass.  Corresponding bands for the other squark and neutralino masses, omitted for clarity of presentation, show similar agreement with the respective observed limits.}
\label{fig:FinLimitsCountNeutralinoAcc}
\end{figure*}

We also compute upper limits on the cross section times branching fraction within the acceptance $A$, where
the latter is defined in the last paragraph of Section~\ref{sec:RecoAndSelection}.
Figures~\ref{fig:FinLimitsCountAccEE} and \ref{fig:FinLimitsCountAccMuMu} show for the electron and muon
channels, respectively, these limits on $\sigma(\Higgsdecay)\BR(\Xdecay)A(\Xdecay)$.
Figure~\ref{fig:FinLimitsCountNeutralinoAcc} shows the corresponding
limits on $\sigma(\sqasq)\BR(\sqdecay)A(\sqdecay)$.  These limits restricted to the acceptance region show
substantially less dependence on the Higgs boson and \X boson masses and on the mean proper decay length $\ctau$ of the \X boson.  They are also less
model dependent, as can be seen by the fact that the limits on $\sigma\BR A$ are similar for \Xdecay and
\chidecay. The residual dependence of the limits on $\ctau$ is due to the $\dxysigma > 12$ requirement at
small values of $\ctau$; whereas at larger values of $\ctau$, it is caused by the fact that, even within the
defined acceptance region, the tracking efficiency falls for leptons produced far from the beamline with
very large impact parameters.

Although the limits described above are determined in the context of two specific models, the analysis is
sensitive to any process in which a LL particle is produced and subsequently decays to a final state that includes dileptons.
To place approximate limits on this more general class of models, one should use
the limits within the acceptance region (\ie on $\sigma\BR A$), because of their smaller model dependence.
In most signal models in which each event contains two identical LL particles that decay in this way, the
limits on $\sigma\BR A$ shown in Figs.~\ref{fig:FinLimitsCountAccEE}--\ref{fig:FinLimitsCountNeutralinoAcc} should
remain approximately valid. (The variation amongst the limit curves shown in these plots for different signal models
and particle masses gives an indication of the accuracy of this statement.)
Exceptions could arise for models that give poor efficiency within the acceptance criteria, \eg for models in
which the leptons are not isolated; have impact parameters with significance below $\dxysigma < 12$,
corresponding to $\abs{d_0}\lesssim 180$\mum; are almost collinear with each other (with the dilepton mass below
15\GeVcc, or for the muon channel $\Delta R < 0.2$); or do not usually satisfy the $\abs{\deltaPhi} < \piOverTwo$ criterion,
such that the parameter $f$ becomes large (\eg if the LL particle is slow-moving and decays to many
particles).

In models where each event contains only one LL particle that can decay inclusively to dileptons, the expected number of selected signal events for given $\sigma\BR$ will be up to a factor of two
lower, and so the limits on $\sigma\BR A$ will be up to a factor of two worse than shown in
Figs.~\ref{fig:FinLimitsCountAccEE}--\ref{fig:FinLimitsCountNeutralinoAcc}.

The acceptance $A$ for any given model can be determined with a generator-level simulation, allowing limits on
$\sigma\BR A$ to be converted to limits on $\sigma\BR$. The following example illustrates this. The limits on
$\sigma(\Higgsdecay)\BR(\Xdecay)$ quoted above are for \Higgs bosons produced through gluon-gluon fusion.  If
the \Higgs bosons were instead produced by the sum of all SM production mechanisms, their momentum spectra
would be slightly harder. For $m_{\Higgs}=125$\GeVcc, the acceptance would then be larger by a factor of
approximately 1.18\,(1.12) for $m_X=20$\,(50)\GeVcc, with a corresponding improvement in the limits on
$\sigma\BR$. The change is smaller for larger \Higgs boson masses.

\section{Summary}

A search has been performed, using proton-proton collision data collected at $\sqrt{s} = 8$\TeV, for long-lived
particles that decay to a final state that includes a pair of electrons or a pair of muons. No such events
have been seen. Quantitative limits have been placed on the product of the cross section and branching fraction
of such a signal in the context of two specific models.  In the first model, a Higgs boson, in the
mass range 125--1000\GeVcc, decays into a pair of hypothetical, long-lived neutral bosons in the mass range 20--350\GeVcc, each of which
can decay to dileptons. The upper limits obtained are typically in the range 0.2--10\fb for
long-lived particles with mean proper decay lengths in the range 0.01--100\cm, and weaken to 2--50\fb for the
lowest considered Higgs mass of 125\GeVcc. In the second model, based on R-parity violating supersymmetry,
a pair of squarks each decays to a quark and a long-lived neutralino \chiz; the neutralino can subsequently decay to $\EE\nu$ or
$\MM\nu$. In this case, the upper limits are typically in the range 0.2--5\fb for \chiz mean proper decay lengths
in the range 0.1--100\cm and squark masses above 350\GeVcc. For a lower squark mass of
120\GeVcc, the limits are typically a factor of ten weaker. These limits are sensitive to branching fractions as
low as $10^{-4}$ and $10^{-6}$ in the Higgs boson and supersymmetric models, respectively.
To allow the results to be reinterpreted in the context of other models, limits that are restricted
to the detector acceptance are also presented, reducing the model dependence.
Over much of the investigated parameter space, these limits are the most stringent to date.

\begin{acknowledgments}
We congratulate our colleagues in the CERN accelerator departments for the excellent performance of the LHC and thank the technical and administrative staffs at CERN and at other CMS institutes for their contributions to the success of the CMS effort. In addition, we gratefully acknowledge the computing centres and personnel of the Worldwide LHC Computing Grid for delivering so effectively the computing infrastructure essential to our analyses. Finally, we acknowledge the enduring support for the construction and operation of the LHC and the CMS detector provided by the following funding agencies: BMWFW and FWF (Austria); FNRS and FWO (Belgium); CNPq, CAPES, FAPERJ, and FAPESP (Brazil); MES (Bulgaria); CERN; CAS, MoST, and NSFC (China); COLCIENCIAS (Colombia); MSES and CSF (Croatia); RPF (Cyprus); MoER, ERC IUT and ERDF (Estonia); Academy of Finland, MEC, and HIP (Finland); CEA and CNRS/IN2P3 (France); BMBF, DFG, and HGF (Germany); GSRT (Greece); OTKA and NIH (Hungary); DAE and DST (India); IPM (Iran); SFI (Ireland); INFN (Italy); MSIP and NRF (Republic of Korea); LAS (Lithuania); MOE and UM (Malaysia); CINVESTAV, CONACYT, SEP, and UASLP-FAI (Mexico); MBIE (New Zealand); PAEC (Pakistan); MSHE and NSC (Poland); FCT (Portugal); JINR (Dubna); MON, RosAtom, RAS and RFBR (Russia); MESTD (Serbia); SEIDI and CPAN (Spain); Swiss Funding Agencies (Switzerland); MST (Taipei); ThEPCenter, IPST, STAR and NSTDA (Thailand); TUBITAK and TAEK (Turkey); NASU and SFFR (Ukraine); STFC (United Kingdom); DOE and NSF (USA).

Individuals have received support from the Marie-Curie programme and the European Research Council and EPLANET (European Union); the Leventis Foundation; the A. P. Sloan Foundation; the Alexander von Humboldt Foundation; the Belgian Federal Science Policy Office; the Fonds pour la Formation \`a la Recherche dans l'Industrie et dans l'Agriculture (FRIA-Belgium); the Agentschap voor Innovatie door Wetenschap en Technologie (IWT-Belgium); the Ministry of Education, Youth and Sports (MEYS) of the Czech Republic; the Council of Science and Industrial Research, India; the HOMING PLUS programme of Foundation for Polish Science, cofinanced from European Union, Regional Development Fund; the Compagnia di San Paolo (Torino); the Consorzio per la Fisica (Trieste); MIUR project 20108T4XTM (Italy); the Thalis and Aristeia programmes cofinanced by EU-ESF and the Greek NSRF; and the National Priorities Research Program by Qatar National Research Fund.
\end{acknowledgments}
\bibliography{auto_generated}

\cleardoublepage \appendix\section{The CMS Collaboration \label{app:collab}}\begin{sloppypar}\hyphenpenalty=5000\widowpenalty=500\clubpenalty=5000\input{EXO-12-037-authorlist.tex}\end{sloppypar}
\end{document}

%% file: EXO-12-037-authorlist.tex
\textbf{Yerevan Physics Institute,  Yerevan,  Armenia}\\*[0pt]
V.~Khachatryan, A.M.~Sirunyan, A.~Tumasyan
\vskip\cmsinstskip
\textbf{Institut f\"{u}r Hochenergiephysik der OeAW,  Wien,  Austria}\\*[0pt]
W.~Adam, T.~Bergauer, M.~Dragicevic, J.~Er\"{o}, M.~Friedl, R.~Fr\"{u}hwirth\cmsAuthorMark{1}, V.M.~Ghete, C.~Hartl, N.~H\"{o}rmann, J.~Hrubec, M.~Jeitler\cmsAuthorMark{1}, W.~Kiesenhofer, V.~Kn\"{u}nz, M.~Krammer\cmsAuthorMark{1}, I.~Kr\"{a}tschmer, D.~Liko, I.~Mikulec, D.~Rabady\cmsAuthorMark{2}, B.~Rahbaran, H.~Rohringer, R.~Sch\"{o}fbeck, J.~Strauss, W.~Treberer-Treberspurg, W.~Waltenberger, C.-E.~Wulz\cmsAuthorMark{1}
% ...
\vskip\cmsinstskip
% ...
1:~~Also at Vienna University of Technology, Vienna, Austria\\
2:~~Also at CERN, European Organization for Nuclear Research, Geneva, Switzerland\\
% ...
